\documentclass[12pt,twoside]{article}

\usepackage{epsf}
\usepackage{times}
\usepackage{a4wide}
\usepackage{epsfig}
\usepackage{fancyheadings,graphics}
\advance \headheight by 3.0truept       

\newlength{\nseparation}
\newenvironment{nfigure}[1]
        {\begin{figure}[#1]\hrule\vspace{\nseparation}\par}
        {\vspace{\nseparation}\par \hrule \end{figure}}

\usepackage{cite}   

\newcommand{\lt}{\left}
\newcommand{\rt}{\right}
\newcommand{\no}{\nonumber}
\newcommand{\nn}{\nonumber \\}
\newcommand{\ov}[1]{\overline{#1}}
\newcommand{\eq}[1]{Eq.~(\ref{#1})}
\newcommand{\eqsand}[2]{Eqs.~(\ref{#1}) and (\ref{#2})}
\newcommand{\eqsto}[2]{Eqs.~(\ref{#1}--\ref{#2})}
\newcommand{\imag}{\mathrm{Im}\,}
\newcommand{\real}{\mathrm{Re}\,}
\newcommand{\gev}{\,\mbox{GeV}}
\newcommand{\mev}{\,\mbox{MeV}}
\newcommand{\sgn}{\mbox{sign}\,}

\newcommand{\Bbar}{\,\overline{\!B}}

\newcommand{\bbd}{\ensuremath{B_d\!-\!\Bbar{}_d\,}}
\newcommand{\bbs}{\ensuremath{B_s\!-\!\Bbar{}_s\,}}

\newcommand{\bbms}{\bbs\ mixing}
\newcommand{\bbmd}{\bbd\ mixing}
\newcommand{\BsorBsbar}{\raisebox{7.7pt}{$\scriptscriptstyle(\hspace*{8.5pt})$}
  \hspace*{-10.7pt}\!\Bbar_{s}} 
\newcommand{\nuornubar}{
  \,\raisebox{5.5pt}{$\scriptscriptstyle(\hspace*{6.3pt})$}
  \hspace*{-7.8pt}\!\ov{\nu}} 
\newcommand{\bra}[1]{\ensuremath{\langle #1 |}}
\newcommand{\ket}[1]{\ensuremath{| #1 \rangle }}
\newcommand{\fig}[1]{Fig.~\ref{#1}}

\newcommand{\lbar}{\ov{\Lambda}}
\newcommand{\dm}{\ensuremath{\Delta M}}
\newcommand{\dg}{\ensuremath{\Delta \Gamma}}
\newcommand{\epm}[2]{
 \raisebox{-0.5ex}{\shortstack[l]{$\scriptstyle+#1$\\$\scriptstyle-#2$}}}

\begin{document}
\thispagestyle{empty}
\begin{flushright}
TTP06-31\\
hep-ph/0612167\\
December 2006
\end{flushright}
\vspace*{0.8cm}
\boldmath
\centerline{\LARGE\bf Theoretical update of \bbs\ mixing}
\unboldmath
\vspace*{1.5cm}
\centerline{{\sc Alexander Lenz}}
\smallskip
\centerline{\sl Institut f{\"u}r Theoretische Physik -- 
                Universit{\"a}t Regensburg}
\centerline{\sl D-93040 Regensburg, Germany}
\medskip
\centerline{and}
\medskip
\centerline{{\sc Ulrich Nierste}}
\smallskip
\centerline{\sl Institut f{\"u}r Theoretische Teilchenphysik -- 
                Universit{\"a}t Karlsruhe}
\centerline{\sl D-76128 Karlsruhe, Germany}
\vspace*{1cm}
\centerline{\bf Abstract}
\vspace*{0.3cm}
\noindent
We update the theory predictions for the mass difference $\dm_s$, the width
difference $\dg_s$ and the CP asymmetry in flavour-specific decays, $a_{\rm
  fs}^{s}$, for the \bbs\ system. In particular we present a new expression
for the element $\Gamma_{12}^s$ of the decay matrix, which enters the
predictions of $\dg_s$ and $a_{\rm fs}^{s}$.  To this end we introduce a new
operator basis, which reduces the troublesome sizes of the $1/m_b$ and
$\alpha_s$ corrections and diminishes the hadronic uncertainty in
$\dg_s/\dm_s$ considerably.  Logarithms of the charm quark mass are summed to
all orders.  We find $\dg_s/\dm_s= (49.7 \pm 9.4 ) \cdot 10^{-4}$ and $\dg_s
=(f_{B_s}/240\,{\rm MeV})^2 [(0.105 \pm 0.016) \, B \; +\; (0.024 \pm 0.004)
\, \tilde{B}_S' \; -\; 0.027 \pm 0.015] \, \mbox{ps}^{-1}$ in terms of the bag
parameters $B$, $\tilde{B}_S'$ in the NDR scheme and the decay constant
$f_{B_s}$. The improved result for $\Gamma_{12}^s$ also permits the extraction
of the CP-violating \bbms\ phase from $a_{\rm fs}^{s}$ with better accuracy.
We show how the measurements of $\Delta M_s$, $\Delta\Gamma_s$, $a_{\rm
  fs}^{s}$, $A_{\rm CP}^{\rm mix}(B_s\to J/\psi \phi)$ and other observables
can be efficiently combined to constrain new physics. Applying our
new formulae to data from the D\O\ experiment, we find a 2$\sigma$ deviation
of the \bbms\ phase from its Standard Model value. We also briefly update the
theory predictions for the \bbd\ system and find $\dg_d/\dm_d = \lt( 52.6
\epm{11.5}{12.8} \rt) \, \cdot 10^{-4}$ and $a_{\rm fs}^d = \lt(
-4.8\epm{1.0}{1.2} \rt) \, \cdot 10^{-4}$ in the Standard Model.

\vspace*{1cm}


\vfill

\newpage

\thispagestyle{empty}
~\\

\newpage
\setcounter{page}{1}
\pagenumbering{arabic}

\section{Introduction}
Flavour-changing neutral current (FCNC) processes are highly sensitive
to new physics around the TeV scale.  Global fits to the unitarity
triangle show an excellent agreement of $b\to d$ and $s\to d$
transitions with the predictions of the Cabibbo-Kobayashi-Maskawa (CKM)
mechanism \cite{ckm,exput}. Extensions of the Standard Model can contain
sources of flavour-changing transitions beyond the CKM matrix.  Models
without these new sources are termed to respect \emph{minimal flavour
  violation (MFV)}.  Despite of the success of the MFV hypothesis in
$b\to d$ and $s\to d$ transitions there is still sizable room for
non-MFV contribution in $b\to s$ transitions. For instance, an extra
contribution to $b\to s \ov{q} q$, $q=u,d,s$, decay amplitudes with a CP
phase different from $\arg(V_{ts}^* V_{tb})$ can alleviate the $\sim
2.6\sigma$ discrepancy between the measured mixing-induced CP
asymmetries in these $b\to s$ penguin modes and the Standard Model
prediction \cite{ichep06}.  Models of supersymmetric grand unification
can naturally accommodate new contributions to $b \to s$ transitions
\cite{cmm}: right-handed quarks reside in the same quintuplets of SU(5)
as left-handed neutrinos, so that the large atmospheric neutrino mixing
angle could well affect squark-gluino mediated $b\to s$ transitions
\cite{jn}.

Clearly, \bbms\ plays a preeminent role in the search for new physics in
$b\to s$ FCNC's.  \bbs\ oscillations are governed by a Schr\"odinger
equation
\begin{equation}
i \frac{d}{dt}
\left(
\begin{array}{c}
\ket{B_s(t)} \\ \ket{\bar{B}_s (t)}
\end{array}
\right)
=
\left( M^s - \frac{i}{2} \Gamma^s \right)
\left(
\begin{array}{c}
\ket{B_s(t)} \\ \ket{\bar{B}_s (t)} 
\end{array}
\right)\label{sch}
\end{equation} 
with the mass matrix $M^s$ and the decay matrix $\Gamma^s$.  The
physical eigenstates $\ket{B_H}$ and $\ket{B_L}$ with the masses
$M_H,\,M_L$ and the decay rates $\Gamma_H,\,\Gamma_L$ are obtained by
diagonalizing $M^s-i \Gamma^s/2$.  The \bbs\ oscillations in
\eq{sch} involve the three physical quantities $|M_{12}^s|$,
$|\Gamma_{12}^s|$ and the CP phase 
$\phi_s=\arg(-M_{12}^s/\Gamma_{12}^s)$ (see e.g.\ 
\cite{run2}).  The mass and width differences between $B_{L}$ and
$B_{H}$ are related to them as
\begin{eqnarray}
\dm_s &=& M^s_H -M^s_L \; = \; 2\, |M_{12}^s|,
\qquad \dg_s \; =\; \Gamma^s_L-\Gamma^s_H \; =\;
        2\, |\Gamma_{12}^s| \cos \phi_s, \label{dmdg}
\end{eqnarray}
up to numerically irrelevant corrections of order $m_b^2/M_W^2$.  
$\dm_s$ simply equals the frequency of the \bbs\ oscillations.
A third quantity providing independent information on the 
mixing problem in \eq{sch} is
\begin{eqnarray}
a^s_{\rm fs}
     &=&
    \imag \frac{\Gamma_{12}^s}{M_{12}^s}
    \; = \; \frac{|\Gamma_{12}^s|}{|M_{12}^s|} \sin \phi_s
    \; = \; \frac{\dg_s}{\dm_s} \tan \phi_s
 . \label{defafs}
\end{eqnarray}
$a_{\rm fs}^s$ is the CP asymmetry in \emph{flavour-specific} $B_s\to f$
decays, which means that the decays $\Bbar_s \to f$ and $B_s \to \ov{f}$
(with $\ov{f}$ denoting the CP-conjugate final state) are forbidden
\cite{hw}. The standard way to access $a_{\rm fs}^s$ uses $B_s \to
X_s \ell^+ \ov{\nu_\ell}$ decays, which justifies the name
\emph{semileptonic CP asymmetry} for $a_{\rm fs}^s$. (See e.g.\ 
\cite{run2,n} for more details on the phenomenology of \bbs\ mixing.)

It is important to note that new physics can significantly affect
$M_{12}^s$, but not $\Gamma_{12}^s$, which is dominated by the
CKM-favoured $b\to c\ov{c}s$ tree-level decays. Hence all possible
effects of new physics can be parameterised by two real parameters only,
for instance $|M_{12}^s|$ and $\phi_s$. While $|M_{12}^s|$ is directly
related to $\dm_s$, the extraction of $\phi_s$ from either $\dg_s$ or
$a_{\rm fs}^s$ requires an accurate knowledge of $\Gamma_{12}^s$. 

In the Standard Model $M_{12}^s$ and $\Gamma_{12}^s$ are computed from the
box diagrams in \fig{fig:box} and QCD corrections in the desired order.  
\begin{nfigure}{t}
\centerline{\epsffile{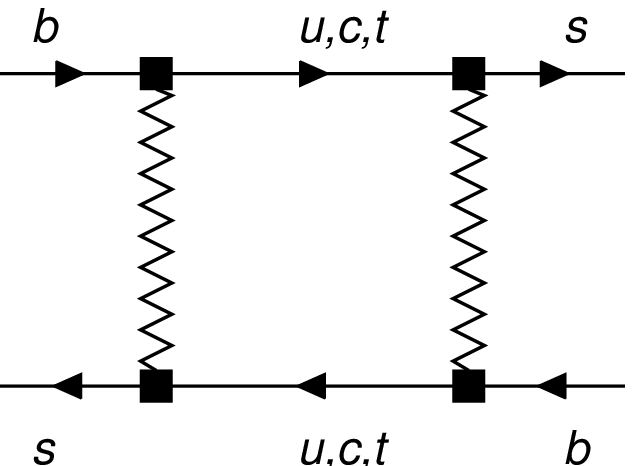}
\hspace{2.5cm}
\parbox[b]{0.25\textwidth}{\epsffile{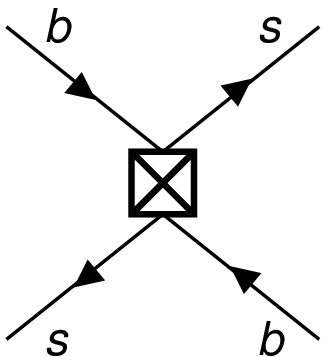}\\[-2mm]}}
\caption{In the lowest order 
  $M_{12}^s$ is calculated from the dispersive parts of the box
  diagrams on the left. It is dominated by the top contribution.
   The result involves only one local 
  $ |\Delta B|=2$ operator, shown in the right picture.  
  The leading contribution to 
  $\Gamma_{12}^s$ is obtained from the absorptive parts of the box
  diagrams on the left, to which only diagrams without top quark line
  contribute. To lowest order in the heavy quark expansion two $ |\Delta B|=2$
  operators occur, the $\lbar/m_b$ corrections involve five more. 
   }\label{fig:box}
\end{nfigure}
The Standard Model prediction for $M_{12}$ reads:
\begin{eqnarray}
M_{12} 
& = &
\frac{G_F^2 M_{B_s} }{12 \pi^2} \, 
M_W^2\, (V_{tb} V_{ts}^*)^2\, \widehat \eta_B \, S_0(x_t) \, 
  f_{B_s}^2 B,
 \label{m12sm}
\end{eqnarray}
where $G_F$ is the Fermi constant, the $V_{ij}$'s are CKM elements, 
$ M_{B_s}$ and $M_W$ are the masses of $B_s$ meson and W boson and
the short-distance information is contained in $ \widehat \eta_B \, S_0(x_t)$:
$S_0 (x_t)$ is the Inami-Lim function, which depends on the top mass
$m_t$ through $x_t=m_t^2/M_W^2$, and $\widehat \eta_B$ is a numerical factor 
containing the leading and next-to-leading QCD corrections \cite{bjw}. 
The calculation of $M_{12}$ involves the four-quark operator
($\alpha,\beta=1,2,3$ are colour indices): 
\begin{eqnarray}
Q & =&   \ov s_\alpha \gamma_\mu (1-\gamma_5) b_\alpha \, 
         \ov s_\beta \gamma^\mu (1-\gamma_5) b_\beta . 
\label{defq} 
\end{eqnarray}
All long-distance QCD effects are contained in the hadronic matrix
element of $Q$ and are parameterised by  $f_{B_s}^2 B$:
\begin{eqnarray}
\bra{B_s} Q \ket{\ov B_s}  &=& \frac{8}{3} M^2_{B_s}\, f^2_{B_s} B 
    .  \label{defb} 
\end{eqnarray}
The recent observation of the \bbs\ mixing frequency $\dm_s=2|M_{12}^s|$
at the Tevatron \cite{dmdiscovery} 
yields a powerful constraint on extensions of the
Standard Model \cite{dmphen, dmphenburas,dmphenligeti,nir2006}. 
The results from the D\O\ and CDF
experiments obtained with 1 fb$^{-1}$ of data, are \cite{ichep06dm}
\begin{eqnarray}
17 \, \mbox{ps}^{-1} \leq  \dm_s & \leq & 21 \, \mbox{ps}^{-1}
  \qquad\qquad \quad @90\% \, \mbox{CL}  \quad\qquad  \mbox{D\O} \nn
\dm_s & =&  17.77\pm{0.10} {}_{\mbox{\scriptsize (syst)}} 
       \pm 0.07\,{}_{\mbox{\scriptsize (stat)}} \, \mbox{ps}^{-1} \, 
   \qquad \mbox{CDF} . \label{dmexp}
\end{eqnarray} 
While the precise measurement in \eq{dmexp} sharply determines
$|M_{12}^s|$, the uncertainty of $f_{B_s}^2 B$, which is around
$30\%$, blurs the extraction of some new physics contribution adding
to $S_0(x_t)$ in \eq{m12sm}. Alternatively one can study the ratio
$\dm_d/\dm_s$, where $\dm_d$ is the mass difference in the \bbd\ system.
While the hadronic uncertainty in the ratio $f_{B_s}^2
B/(f_{B_d}^2 B_{B_d})$ is smaller, one is now dependent on
$|V_{td}/V_{ts}|^2$. Even if one assumes non-standard contributions
only in $B_s$ physics, but not in the quantities entering the global fit
of the unitarity triangle, $|V_{td}/V_{ts}|^2$ is only known to roughly
$40\% $ \cite{exput} leaving equally much room for new physics in
$|M_{12}^s|$.

Adding experimental information from $\dg_s$ or $a_{\rm fs}^s$ helps in
two ways; first, one can study the CP-violating phase $\phi_s$, which is
totally unconstrained by $\dm_s$, through \eqsand{dmdg}{defafs}.
Second, one expects cancellations of hadronic parameters in the ratio
$\Gamma_{12}^s/M_{12}^s$, which enters $a_{\rm fs}^s$ and $\dg_s/\dm_s$.
All decays into final states with zero strangeness
contribute to $\Gamma_{12}^s$, which is dominated by the   
CKM-favoured $b\to c\ov{c}s$ tree-level contribution. 
In the first step of the calculation the W-boson is integrated out and 
the W-mediated $|\Delta B|=1$ transitions are described by the usual 
effective $|\Delta B|=1$ hamiltonian with the current-current operators 
$Q_1$, $Q_2$ and the penguin operators $Q_{3-6}$, $Q_8$ \cite{bjlw}. 
The leading contribution to  $\Gamma_{12}^s$ in this effective  
$|\Delta B|=1$ theory is shown in \fig{fig:dga}. 
\begin{nfigure}{t}
\centerline{\epsffile{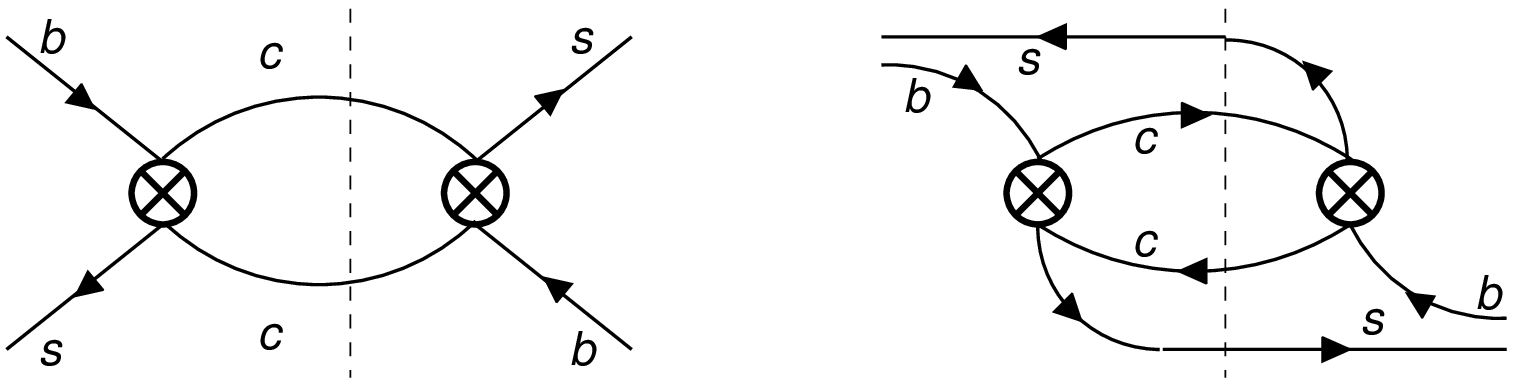}}
\caption{Leading-order CKM-favoured contribution to $\Gamma_{12}^s$, 
  which arises 
  from $\protect\BsorBsbar$ decays to final states (indicated by the dashed
  lines) with a $(c,\ov{c})$ pair and zero strangeness.  The crosses denote
  any of the operators $Q_{1-6}$ of the $|\Delta B| =1$ hamiltonian.
  The Cabibbo-suppressed contributions correspond to diagrams with one or both
  $c$ quarks replaced by $u$ quarks.}\label{fig:dga}
\end{nfigure}
In the second step one uses an operator product expansion (OPE), the
Heavy Quark Expansion (HQE), to express $\Gamma_{12}^s$ as an expansion
in the two parameters $\lbar/m_b$ and $\alpha_s(m_b)$.  Here $\alpha_s$
is the QCD coupling constant and $\lbar$ is the appropriate hadronic
scale, which quantifies the size of the hadronic matrix elements. The
HQE links the diagrams of \fig{fig:dga} to the matrix elements of local
$\Delta B =2$ operators. In addition to the operator $Q$ in \eq{defq}
one also encounters
\begin{eqnarray}
Q_S & = &  \ov{s}_\alpha (1+\gamma_5)  b_\alpha \, 
           \ov{s}_\beta  (1+\gamma_5)  b_\beta ,
  \label{defqs}
\end{eqnarray}
whose matrix element is parameterised by a bag parameter $B_{S}$ in
analogy to \eq{defb}. The leading contribution to $\Gamma_{12}^s$ was
obtained in \cite{hw,LO}. Today  $\Gamma_{12}^s$ is known to 
next-to-leading-order (NLO) in both $\lbar/m_b$ \cite{bbd1} and 
$\alpha_s(m_b)$ \cite{bbgln1,rome03}. The 1998 result \cite{bbgln1}
\begin{equation}
\left(\frac{\dg_s}{\Gamma_s}\right) =
\left(\frac{f_{B_s}}{210~{\rm MeV}}\right)^2
\left[0.006\, B + 0.150\, B_S  - 0.063\right]
\label{olddgg}
\end{equation}
with the average total width $\Gamma_s=(\Gamma^s_L+\Gamma^s_H)/2$ is
pathological in several respects: first, the $\lbar/m_b$ correction -0.063 is
unnaturally large and amounts to around 40\% of the total result. Second, the
coefficient of $B$ cancels almost completely, the result is therefore
dominated by the term proportional to $B_S\sim 0.9$, so that the cancellation
of hadronic quantities from the ratio $\dg_s/\dm_s$ is very imperfect. Third,
both the $\lbar/m_b$ and $\alpha_s$ corrections, which diminish the
coefficient of $B_S$ from 0.22 to 0.15, are negative, and these numerical
cancellations between leading-order (LO) order result and corrections increase
the relative uncertainty of the prediction for $\dg_s/\Gamma_s$.  In the
following section we argue that these pathologies are caused by a poor choice
of the operator basis used in \cite{bbd1,bbgln1,rome03} and propose a
different basis. We also improve the prediction of $\dg_s/\dm_s$ and
$\dg_s/\Gamma_s$ in several other aspects, by summing logarithms of the charm
mass to all orders in $\alpha_s$, by using different renormalisation schemes
for the $b$-quark mass, by including CKM-suppressed contributions and by
modifying the normalisation related to the factor $1/\Gamma_s$ in \eq{olddgg}.
In Sect.~\ref{sect:num} we present numerical updates first of $\dm_s$,
$\dg_s$ and $a_{\rm fs}^s$ and then of the corresponding quantities in the
$B_d$-system.
In Sect.~\ref{sect:new} we show how the expressions for the mixing quantities
change in the presence of new physics. Here we discuss how to combine
different present and future measurements to constrain $|M_{12}^s|$ and
$\phi_s$ and advocate a novel method to display the constraints on possible
new short-distance physics in \bbs\ mixing.  Sect.~\ref{sect:road} gives a
road map for future measurements and calculations and Sect.~\ref{sect:sum}
summarises our results.  

\section{Improved prediction of $\Gamma_{12}^s$}
We write $\Gamma_{12}^s$ as \cite{bbln}
\begin{eqnarray}
\Gamma_{12}^s &=& - \lt[\, \lambda_c^2 \, \Gamma_{12}^{cc} \; + \;
        2 \, \lambda_c\,  \lambda_u \, \Gamma_{12}^{uc}\; + \;
        \lambda_u^{2} \, \Gamma_{12}^{uu} \,  \rt] \label{ga12} \\
      &=& 
    - \lt[\, \lambda_t^2 \, \Gamma_{12}^{cc} \; + \;
        2 \, \lambda_t\,  \lambda_u \, 
         \lt( \Gamma_{12}^{cc} - \Gamma_{12}^{uc} \rt) \; + \;
        \lambda_u^{2} \, 
        \lt( \Gamma_{12}^{cc} - 2 \Gamma_{12}^{uc} + \Gamma_{12}^{uu} \rt)  
        \rt] \label{ga12t} 
\end{eqnarray}
with the CKM factors $\lambda_i=V_{is}^* V_{ib}$ for $i=u,c,t$.  In \eq{ga12t}
we have eliminated $\lambda_c$ in favour of $\lambda_t$ using 
$\lambda_u+\lambda_c+\lambda_t=0$ to prepare for the study of
 $\Gamma_{12}^s/M_{12}^s$. Since $|\lambda_u|\ll |\lambda_t|\approx
|\lambda_c|$, $\Gamma_{12}^{cc}$ clearly dominates 
 $\Gamma_{12}^s$. For $ab=cc,uc,uu$ we write \cite{bbgln1,bbln}
\begin{eqnarray}
\Gamma_{12}^{ab} &=& \frac{G_F^2 m_b^2}{24 \pi\, M_{B_s}} 
    \lt[ \, G^{ab} \, \bra{B_s } Q \ket{\Bbar_s} \; - \;
          G_S^{ab} \,\bra{B_s} Q_S \ket{\Bbar_s} 
    \rt] \; +\; \Gamma_{12,1/m_b}^{ab} \label{defg}
\end{eqnarray}
The coefficients $G^{ab}$ and $G_S^{ab}$ are further decomposed as 
\begin{eqnarray}
G^{ab} \;=\; F^{ab} + P^{ab}, \qquad  
G_S^{ab} \; = \; - F_S^{ab} -  P_S^{ab} 
    . \label{deffp}
\end{eqnarray}
Here $F^{ab}$ and $F_S^{ab}$ are the contributions from the current-current
operators $Q_{1,2}$ while the small coefficients $P^{ab}$ and $P_S^{ab}$ stem
from the penguin operators $Q_{3-6}$ and $Q_8$.  (Note that in
\cite{bbgln1}, where only the dominant $\Gamma_{12}^{cc}$ was considered,
these coefficients had no superscript 'cc'.)  Numerical cancellations render
$F^{cc}$ small with $|F^{cc}/F_S^{cc}|\approx 0.03$ which explains the small
coefficient of $B$ in \eq{olddgg}.

We parameterise the matrix element of $Q_S$ as
\begin{eqnarray}
\bra{B_s} Q_S \ket{\Bbar_s} &=& -\frac{5}{3}  M^2_{B_s}\,
                  f^2_{B_s} B_S^\prime . \label{defbsp} 
\end{eqnarray}
Formulae for physical quantities are more compact when expressed in terms of
$B_S^\prime$ rather than the conventionally used bag parameter
$B_S$. The two parameters are related as
\begin{eqnarray}
B_S^\prime &=& 
  \frac{M^2_{B_s}}{(\ov m_b+\ov m_s)^2} B_S
\label{defbs} .
\end{eqnarray}
In the vacuum insertion approximation (VIA) the bag factors $B$ and
$B_S$ are equal to one. Throughout this paper we use the $\ov{\rm MS}$
scheme  as defined in \cite{bbgln1,bbln} for all operators. Therefore the
masses $\ov m_b$ and $\ov m_s$ appearing in \eq{defbs} correspond to the
$\ov{\rm MS}$ scheme as well.

$\Gamma_{12,1/m_b}^{cc}$ comprises effects suppressed by $\lbar/m_b$.
We will discuss it later, after transforming to our new operator basis.   

\subsection{New operator basis}
When calculating $\Gamma_{12}$ to leading order in $\lbar/m_b$, one first
encounters a third operator $\widetilde Q_S$ in addition to $Q$ and $Q_S$
defined in \eqsand{defq}{defqs}:
\begin{eqnarray}
\widetilde{Q}_S & = &  \ov{s}_\alpha (1+\gamma_5)  b_\beta \, 
                       \ov{s}_\beta (1+\gamma_5)  b_\alpha,
  \label{defqst}
\end{eqnarray}
However, a certain linear combination of $Q$, $Q_S$ and 
$\widetilde Q_S$ is a $1/m_b$--suppressed operator \cite{bbd1}. 
This $1/m_b$--suppressed operator reads
\begin{eqnarray}  
R_0 &\equiv&                            Q_S 
              \; + \; \alpha_1  \tilde Q_S 
           \; + \; \frac{1}{2}  \alpha_2 Q, 
\label{defr0}
\end{eqnarray}  
where  $\alpha_{1,2}$ contain NLO corrections, which are 
specific to the $\ov{\rm MS}$ scheme used by us \cite{bbgln1}:
\begin{eqnarray}
 \!  \!  \alpha_1 \;=\; 1 \; +\;  \frac{\alpha_s(\mu_2)}{4\pi} \,
            C_f \lt( 12 \ln \frac{\mu_2}{m_b} + 6 \rt),    
       &&\; 
  \alpha_2 \;=\; 1 \; +\; \frac{\alpha_s(\mu_2)}{4\pi} \, 
           C_f \lt( 6 \ln \frac{\mu_2}{m_b} + \frac{13}{2} \rt).        
\label{defr12}
\end{eqnarray}  
Here $C_f=4/3$ is a colour factor and $\mu_2$ is the scale at which the
operators in \eq{defr0} are defined. The coefficients $G$ and $G_S$ in
\eq{defg} depend on $\mu_2$ and this dependence cancels with the
$\mu_2$--dependence of $\bra{B_s} Q (\mu_2)\ket{\ov B_s}$ and $\bra{
  B_s} Q_S(\mu_2) \ket{\ov B_s}$. In lattice computations the
$\mu_2$--dependence enters in the lattice--continuum matching of these
matrix elements. In our numerics we will always quote the results for
$\mu_2=m_b$.  In \cite{bbd1,bbgln1,rome03} \eq{defr0} has been used to
eliminate $Q_S$ in favour of $R_0$ leading to the result in \eq{olddgg}.
The matrix element of $\widetilde Q_S$ reads
\begin{eqnarray}
\bra{B_s} \widetilde Q_S (\mu_2)\ket{\ov B_s} &=& \frac{1}{3}  M^2_{B_s}\,
                  f^2_{B_s} \widetilde B_S^\prime (\mu_2). 
    \label{defbstp} 
\end{eqnarray}
In analogy to \eq{defbs} we define 
\begin{eqnarray}
\widetilde B_S^\prime (\mu_2) &=& 
  \frac{M^2_{B_s}}{(\ov m_b (\mu_2)+\ov m_s(\mu_2))^2} 
  \widetilde B_S (\mu_2)
\label{defbst} .
\end{eqnarray}
For clarity we have explicitly shown the $\mu_2$-dependence in
\eqsand{defbstp}{defbst}, which was skipped in
Eqs.~(\ref{defb}),(\ref{defbsp}) and (\ref{defbs}).  In VIA $\widetilde
B_S =1$ and $\bra{B_s} \widetilde Q_S \ket{\ov B_s}$ is much smaller
than $\bra{B_s} Q \ket{\ov B_s}$ and $\bra{B_s} Q_S \ket{\ov B_s}$. The
small coefficient $1/3$ in \eq{defbstp} is the consequence of a
cancellation between the leading term in the $1/N_c$ expansion, where
$N_c=3$ is the number of colours, and the factorisable $1/N_c$
corrections: $1/3=1-2/N_c$. One naturally expects that the bag factor
$\widetilde B_S$ substantially deviates from 1.  However, a lattice
computation found $\widetilde B_S = 0.91 \pm 0.08$ \cite{bgmpr}, showing
that the matrix element of $\widetilde Q_S$ is indeed small. Thus
$\bra{B_s} R_0 \ket{\ov B_s}=\lbar/m_b$ implies a strong numerical
relationship between $B$ and $B_S$ which can be used to constrain
$B_S/B$ entering $\dg_s/\dm_s$. Yet it is more straightforward to use
\eq{defr0} to eliminate $Q_S$ altogether from $\Gamma_{12}$ in favour of
$\widetilde Q_S$. The coefficient of $B$ will change and and the
coefficient of $\widetilde B_S^\prime$ is expected to be small in view
of the factor of 1/3 in \eq{defbstp}. Using further the bag parameters
of \eqsand{defb}{defbsp}, $\Gamma_{12}^{ab}$ of \eq{defg} now reads
\begin{eqnarray}
\Gamma_{12}^{ab} &=& \frac{G_F^2 m_b^2}{24 \pi} \, M_{B_s}\,
                  f^2_{B_s} \, 
    \lt[ \, \lt( G^{ab} + \frac{\alpha_2}{2} G_S^{ab} \rt) \, 
            \frac{8}{3} B \; + \;
              G_S^{ab} \alpha_1 \, \frac{1}{3} 
              \widetilde B_S^\prime
    \rt] \; +\; \widetilde \Gamma_{12,1/m_b}^{ab} .\label{g12n}
\end{eqnarray}
The new $1/m_b$--corrections are related to $\Gamma_{12,1/m_b}^{ab}$ appearing
in \eq{defg} as
\begin{eqnarray}
    \widetilde \Gamma_{12,1/m_b}^{ab} &=& 
      \Gamma_{12,1/m_b}^{ab} \; +\; 
     \frac{G_F^2 m_b^2}{24 \pi\, M_{B_s}} 
     \, F_S^{ab,(0)} \, \bra{B_s} R_0 \ket{\ov B_s} 
   .\label{g12t}
\end{eqnarray}
Here we have taken into account that the result of \cite{bbgln1,rome03}
includes the $\lbar/m_b$ terms without penguin contributions and to LO
in $\alpha_s$: consequently we have changed $-G_S^{ab}$ to
$F_S^{ab,(0)}$, which is the LO approximation to $F_S^{ab}$. Recalling
$|G^{ab}|\ll |G_S^{ab}|$ and $B,\widetilde B_S^\prime
\approx 1$ one easily verifies from \eq{g12n} that the first term
proportional to $B$ dominates over the second term. Since
$\Gamma_{12,1/m_b}^{ab}$ in \eq{olddgg} is negative and the shift in
\eq{g12t} adds a positive term our change of basis also leads to
$|\widetilde \Gamma_{12,1/m_b}^{ab}|<|\Gamma_{12,1/m_b}^{ab}|$. Further
the $\alpha_s$-corrections contained in $\alpha_{1,2}$, which multiply
$G_S^{ab,(0)}$ in \eq{g12n}, temper the large NLO corrections of the old
result.  These three effects combine to reduce the hadronic uncertainty
in $\dg_s/\dm_s$ substantially. In other words: the uncertainty quoted
in \cite{bbgln1,rome03} is not intrinsic to $\dg_s/\dm_s$ but an
artifact of a poorly chosen operator basis.

\subsection{A closer look at $\mathbf{1/m_b}$ corrections}
At order $1/m_b$ one encounters the operators $R_0$ of \eq{defr0},  
\begin{eqnarray}
R_1 & = & \frac{m_s}{m_b} \;  
\ov s_\alpha (1+\gamma_5) b_\alpha \, \ov s_\beta (1-\gamma_5)b_\beta \nn
R_2 & = & \frac{1}{m^2_b} \;
\ov s_\alpha {\overleftarrow D}_{\!\rho} \gamma^\mu (1-\gamma_5) 
    D^\rho b_\alpha\, 
\ov s_\beta \gamma_\mu (1-\gamma_5) b_\beta  \nn
R_3 &= & 
\frac{1}{m^2_b} \; 
\ov s_\alpha {\overleftarrow D}_{\!\rho} (1+\gamma_5) D^\rho b_\alpha \, 
\ov s_\beta (1+\gamma_5) b_\beta  
\label{defr}
\end{eqnarray}
and the operators $\widetilde{R}_i$ which are obtained from the $R_i$'s
by interchanging the colour indices $\alpha$ and $\beta$ of the two $s$
fields \cite{bbd1}. At order $1/m_b$ only five of these operators are 
independent because of relations like 
$\widetilde{R}_2=-R_2+{\cal O}(1/m_b^2)$.  
Writing (for $ab=cc,uc,uu$)
\begin{eqnarray}
 \widetilde{\Gamma}_{12,1/m_b}^{ab} 
&=&  \frac{G_F^2 m_b^2}{24 \pi M_{B_s}} 
   \lt[ g_0^{ab} \bra{B_s} R_0 \ket{\ov B_s}  \, + \,
   \sum_{j=1}^3 \lt[ g_j^{ab} \bra{B_s} R_j \ket{\ov B_s} + 
       \widetilde{g}_j^{ab} \bra{B_s} \widetilde{R}_j \ket{\ov B_s} 
       \rt]
      \rt] \label{ga12m}
\end{eqnarray}
the coefficients $g_j^{ab}$ and $\widetilde{g}_j^{ab}$ 
read \cite{bbd1,dhky,bbln}:
\begin{eqnarray}
g_0^{cc} \; =\; \sqrt{1-4 z}  (1+2z) C_2^{(0)\, 2} \, + \,
        F_S^{cc\,(0)}  \; =  &&\!\!\!\!\!
   \sqrt{1-4 z} \,  (1+2z) \, C_1^{(0)}
  \lt[ 3  C_1^{(0)} \, +\, 2\,  C_2^{(0)}  \rt]
 \nn
g_1^{cc} \; =\; -2 \sqrt{1-4 z} (1+2z) \, C_1^{(0)} \,
           \lt[ 3 C_1^{(0)}\, + \, 2 C_2^{(0)} \rt]
&&\quad
\widetilde{g}_1^{cc} \; =\; -2 \sqrt{1-4 z} (1+2z) C_2^{(0)\,2} \nn
g_2^{cc} \; =\; -2 \, \frac{1-2z-2z^2}{\sqrt{1-4 z}}  \, C_1^{(0)} \,
     \lt[ 3 C_1^{(0)}\, + \, 2 C_2^{(0)} \rt]
&&\quad
\widetilde{g}_2^{cc} \; =\;
   - 2 \, \frac{1-2z-2z^2}{\sqrt{1-4 z}} C_2^{(0)\,2} \nn
g_3^{cc}  \; =\; -24\,  \frac{z^2}{\sqrt{1-4 z}} \, C_1^{(0)}
             \lt[ 3 C_1^{(0)}\, + \, 2 C_2^{(0)} \rt]
&&\quad
\widetilde{g}_3^{cc} \; =\; -24\,
         \frac{z^2}{\sqrt{1-4 z}} C_2^{(0)\, 2} \label{gcc}
\end{eqnarray}
\begin{eqnarray}
g_0^{uc} \; =\; (1-z)^2 (1+2z) C_2^{(0)\, 2} \, + \,
        F_S^{uc\,(0)}  \, =  &&\!\!\!\!\!\!\!
   (1-z)^2 \,  (1+2z) \, C_1^{(0)}
  \lt[ 3  C_1^{(0)} \, +\, 2\,  C_2^{(0)}  \rt]
 \nn
g_1^{uc} \; =\; -2 (1-z)^2 (1+2z) \, C_1^{(0)} \,
           \lt[ 3 C_1^{(0)}\, + \, 2 C_2^{(0)} \rt]
&&
\widetilde{g}_1^{uc} \; =\; -2 (1-z)^2 (1+2z) C_2^{(0)\,2} \nn
g_2^{uc} \; =\; - 2 \, (1-z)\, (1+z+z^2)  \, C_1^{(0)} \,
     \lt[ 3 C_1^{(0)}\, + \, 2 C_2^{(0)} \rt]
&&
\widetilde{g}_2^{uc} \; =\;
   - 2 \, (1-z)\, (1+z+z^2) C_2^{(0)\,2} \nn
g_3^{uc}  \; = \; - 12 \, (1-z) \, z^2 \, C_1^{(0)}
             \lt[ 3 C_1^{(0)}\, + \, 2 C_2^{(0)} \rt]
&&
\widetilde{g}_3^{uc} \; =\; - 12 \, (1-z) \,
         z^2 \, C_2^{(0)\, 2} . \label{guc}
\end{eqnarray}
and $g_j^{uu}=g_j^{cc}(z=0)=g_j^{uc}(z=0)$.  Here 
\begin{eqnarray}
z &\equiv& \frac{\ov m_c^2}{\ov m_b^2} \; \equiv\; 
  \frac{\lt[ \ov m_c(\ov m_c)\rt]^2}{ \lt[ \ov m_b(\ov m_b)\rt]^2 } 
      \label{defz}
\end{eqnarray}
and $C_1^{(0)}\sim -0.3$ and $C_2^{(0)}\sim 1.1$ are the LO Wilson
coefficients of the $\Delta B =1$ operators $Q_1$ and $Q_2$ \cite{bjlw}.

The contributions involving $R_1$, $\widetilde R_1$, $R_3$ and
$\widetilde R_3$ are suppressed by powers of $m_s/m_b$ or $z^2$ and are
numerically negligible. The only two important $1/m_b$ operators are
$R_0$ and $\widetilde{R}_2=-R_2+{\cal O}(1/m_b^2)$.  As a consequence of
the elimination of $Q_S$ in favour of $\widetilde{Q}_S$ no term
involving the large coefficient $C_2^{(0)\, 2} $ occurs in $g_0^{ab}$.
The contribution from $R_0$ is substantially diminished, and this can be
understood in terms of a systematic expansion in $1/N_c$: the
coefficients $g_0^{ab}$ are colour--suppressed due to $C_1\sim 1/N_c$,
while they were colour--favoured in the old basis. Since radiative
corrections cannot change the colour counting, this feature must persist
in the yet uncalculated order $\alpha_s/m_b$. In other words, by
changing to our new basis we have absorbed the corrections of order
$N_c^0/m_b$ into the leading order of the $1/m_b$ expansion. This
improves our result over the one in the old basis by a term of order
$N_c \alpha_s/m_b$. (Recall that $\alpha_s\sim 1/N_c$, so that $N_c
\alpha_s/m_b \sim N_c^0/m_b$.)  This term (which constitutes a
parametrically enhanced correction) would appear, if the calculation of
$\alpha_s/m_b$ were done in the old basis. In fact, this term occurs in
the NLO calculation of \cite{bbgln1,rome03,bbln} in the coefficient of
$\widetilde{Q}_S$ but is dropped once $\widetilde{Q}_S$ is traded for
$R_0$, because all $\alpha_s/m_b$ terms are consistently discarded.
With the use of our new basis no corrections of order $N_c \alpha_s/m_b$
to $g_0^{ab}$ can occur.  This feature can also be understood by
realising that the large--$N_c$ contribution to $\Gamma_{12}^{ab}$ stems
from the right diagram in \fig{fig:dga} with two insertions of $Q_2$
plus additional planar graphs with extra gluons. These diagrams
contribute to the coefficients of $Q$ and $\widetilde{Q}_S$, but not to
the coefficient of $Q_S$. (This is easy to see, if one inserts the two
$Q_2$'s in the Fierz-rearranged form.)  Upon elimination of $Q_S$ in
favour of $R_0$, the color--suppressed coefficient $g^{ab}$ of $Q_S$
becomes the coefficient of $R_0$.  At order $1/m_b$ one has to include
the momentum of the $s$ quark in that diagram and finds a contribution
to the $\widetilde{g}_i^{ab}$'s at order $N_c^0$.  These terms are
identical in both bases. Our numerical analysis in Sect.~\ref{sect:num}
follows the pattern revealed by the $1/N_c$ expansion, finding the
numerical relevance of $R_0$ drastically reduced compared to the old
basis, so that the only remaining important $1/m_b$ operator is
$\widetilde R_2$.
 
In the new basis the $1/m_b$ corrections have their natural size of
order $\lbar/m_b\sim 20\%$. To be conservative, we have estimated the
$1/m_b^2$ terms to verify that this result is not accidental. We have
found two types of contributions: the first type is calculated by
expanding the results of \fig{fig:dga} to the next order of the
$s$--quark momentum, yielding operators with more derivatives acting on
the $s$ quark field. We find that these contributions have the same
suppression pattern as the $g_i^{ab}$'s and $\widetilde{g}_i^{ab}$'s.
The second type of $1/m_b^2$ operators involve the QCD field strength
tensor $G_{\mu\nu}$ and has no counterparts at lower orders. We find
small coefficients here as well. Since the size of the $1/m_b^2$
corrections is well below the uncertainty which we obtain by varying the
bag factors of the operators in \eq{defr}, there is no reason to include
these corrections into our numerical code.

We parameterise the matrix elements $\langle R_i\rangle \equiv \bra{B_s}
R_i \ket{\ov B_s} $'s as
\begin{eqnarray}
\lefteqn{\hspace{-3cm}\langle  R_0 \rangle \; =\;  - \, \frac{4}{3} 
   \lt[ \frac{M_{B_s}^2}{m_b^{\rm pow\, 2} 
   \lt( 1 + \ov{m}_s/\ov{m}_b \rt)^2 } - 1 \rt] 
  M_{B_s}^2 f_{B_s}^2 B_{R_0},} \nn 
 \langle  R_1 \rangle \; =\; \frac{7}{3} \,
   \frac{\ov{m}_s}{\ov{m}_b} M_{B_s}^2 f_{B_s}^2 B_{R_1}, && 
 \langle  \widetilde{R}_1 \rangle \; =\; \frac{5}{3} \,
   \frac{\ov{m}_s}{\ov{m}_b} M_{B_s}^2 f_{B_s}^2 B_{\widetilde{R}_1}, \nn 
 \langle  R_2 \rangle \; =\; -\,  \frac{2}{3} \, 
    \lt[ \frac{M_{B_s}^2}{m_b^{\rm pow\, 2}} -1 \rt] \,  
    M_{B_s}^2 f_{B_s}^2 B_{R_2}, && 
 \langle  \widetilde{R}_2 \rangle \; =\; \frac{2}{3} 
    \lt[ \frac{M_{B_s}^2}{m_b^{\rm pow\, 2}} -1 \rt] \,  
    M_{B_s}^2 f_{B_s}^2 B_{\widetilde{R}_2}, \nn 
 \langle  R_3 \rangle \; =\; \frac{7}{6} \, 
    \lt[ \frac{M_{B_s}^2}{m_b^{\rm pow\, 2}} -1 \rt] \,  
    M_{B_s}^2 f_{B_s}^2 B_{R_3}, && 
 \langle  \widetilde{R}_3 \rangle \; =\; \frac{5}{6} 
    \lt[ \frac{M_{B_s}^2}{m_b^{\rm pow\, 2}} -1 \rt] \,  
    M_{B_s}^2 f_{B_s}^2 B_{\widetilde{R}_3}. \label{melr}
\end{eqnarray}
As usual the bag parameters $B_{R_0},\ldots,B_{\widetilde{R}_3}$
parameterise the deviation of the matrix elements from their VIA results
derived in \cite{bbd1}. The numerical values of the $\langle
R_i\rangle$'s depend sensitively on the choice of the mass parameter
$m_b^{\rm pow}$ in \eq{melr}. Clearly, $m_b^{\rm pow}$ is a redundant
parameter, as any change in $m_b^{\rm pow}$ can be absorbed into the bag
parameters. It merely serves to calibrate the overall size of the
$1/m_b$-suppressed matrix elements such that the bag factors are close
to 1. A future NLO calculation of the coefficients in \eq{guc} will
allow us to replace $m_b^{\rm pow}$ by a well-defined (i.e.\ properly
infrared-subtracted) $b$ pole mass.  Our numerical value for $m_b^{\rm
  pow}$ is guided by the requirement that the terms in square brackets
in \eq{melr} are of order $2 \lbar/ m_b^{\rm pow}\sim 0.2$, which leads
to the estimate $m_b^{\rm pow}\approx 4.8 \, \gev$. A better
justification can be given by noting that the lattice computations of
$B$, $B_S$ and $\widetilde{B}_S$ in \cite{bgmpr} allow for an estimate
of $\langle R_0 \rangle$ (which may become a determination, once the
lattice-continuum matching of $\langle R_0 \rangle $ is done at NLO):
\begin{eqnarray}
B_{R_0} &=& \lt[ \frac{\alpha_1}{4} \widetilde{B}_S^\prime + \alpha_2 B -
  \frac{5}{4} B_S^\prime   \rt]
     \lt[ 1 - \frac{M_{B_s}^2}{m_b^{\rm pow\,2} 
        (1+\ov{m}_s/\ov{m}_b)^2} \rt]^{-1} \label{brn}
\end{eqnarray}
With the central values for $B$, $B_S$ and $\widetilde{B}_S$ given in
\cite{bgmpr} and the choice $m_b^{\rm pow} = 4.8 \, \gev$ one finds
$B_{R_0}=1.1$, while those of the new preliminary lattice computation of
\cite{latticeall} imply $B_{R_0}=1.7$. Our quoted numerical results in
Sect.~\ref{sect:num} correspond to conservative ranges for both
$m_b^{\rm pow}$ and the $B_{R_i}$'s. We note that the only places where
we use $m_b^{\rm pow}$ are the matrix elements in \eq{melr}; it is not
used in the overall factor $m_b^2$ of
$\widetilde{\Gamma}_{12,1/m_b}^{ab}$ in \eq{ga12m}. This is a change
compared to the analysis in \cite{bbln}.

\boldmath
\subsection{Summing terms of order $\alpha_s^n z \ln^n z$}
\unboldmath%
The coefficients $G^{ab}$ and $G_S^{ab}$ in \eq{defq} depend on quark
masses through $z$ defined in \eq{defz}. At order $\alpha_s^n$ the
dominant $z$-dependent terms are of the form $\alpha_s^n z \ln^n z$.  In
\cite{BBGLN02} and \cite{bbln} it has been shown that these terms are
summed to all orders $n=1,2,\ldots$, if one switches to a
renormalisation scheme which uses
\begin{eqnarray}
\ov z &\equiv& 
  \frac{\lt[ \ov m_c(\ov m_b)\rt]^2}{ \lt[ \ov m_b(\ov m_b)\rt]^2 } 
   .     \label{defzb}
\end{eqnarray}
Since $\ov z$ is roughly half as big as $z$, this also reduces the 
dependence of the coefficients on the charm mass. We illustrate the
effect for $\Gamma_{12}^{cc}$ with a numerical example:   
In the two renormalisation schemes one finds
\begin{eqnarray}
\Gamma_{12}^{cc} &=&  \lt( 3.3 \,-\, 11.4\, z \,+\, 
                        1.5  \, z \, \ln z \rt) \cdot 10^{-3} 
    \, \mbox{ps}^{-1}   
        \; + \;{\cal O}
  \lt( z^2 \rt) \nn
\Gamma_{12}^{cc} &=&  \lt( 3.3 \,-\, 11.4 \, \ov z \rt) \cdot 10^{-3}  
    \, \mbox{ps}^{-1} \; +\;  {\cal O}
  \lt( \ov z^2 \rt) . \label{illu} 
\end{eqnarray}
The numerical input is taken from \eqsto{mb}{mw} and \eq{setone} below.  From
\eq{illu} one verifies that the use of $\ov z$ eliminates the $z \ln z$ term.
This issue is particularly relevant for $a^s_{\rm fs}$ and $a^d_{\rm fs}$,
which are of order $z$.  The final numbers for all quantities quoted below
involve $\ov z$. We only revert to a scheme using $z$ to compare with the
previously published results in \cite{bbgln1,rome03}.

\section{Numerical predictions}\label{sect:num}
\subsection{Input}
For the numerical analysis we use the following set of input parameters:
The quark masses are \cite{bmass}
\begin{eqnarray}
\label{mb}
\ov m_b( \ov m_b) & =& 4.22 \pm 0.08 \, {\rm GeV}
\qquad \Rightarrow \quad 
m_b^{\rm pole} = 4.63 \pm 0.09 ~{\rm GeV}\,   \\
m_b^{\rm pow} &=&  4.8^{+0.0}_{- 0.2}  \, {\rm GeV}\, \nn  
\label{mc}
\ov m_c( \ov m_c) & =&  1.30 \pm 0.05 \, {\rm GeV}
\; \Rightarrow \;
z = \frac{\ov m_c^2 (\ov m_c)}{\ov m_b^2 (\ov m_b)} = 0.095 \pm  0.008 \, ,
\\
&& \phantom{1.30 \pm 0.05 \, {\rm GeV}\;}
\Rightarrow \;
\ov z = \frac{\ov m_c^2(\ov m_b)}{\ov m_b^2(\ov m_b)} = 
0.048 \pm 0.004\, \no \\[2mm] 
\label{ms}\ov m_s(2 \, {\rm GeV}) & =&  0.10 \pm 0.02 \, {\rm GeV} 
\quad \Rightarrow\quad
\ov m_s(\ov m_b) = 0.085 \pm 0.017 \, {\rm GeV}\, \nn 
\label{mt} 
m_t^{\rm pole} &=& 171.4 \pm 2.1 \,  {\rm GeV} \quad \Rightarrow\quad  
\ov m_t (\ov m_t) = 163.8 \pm 2.0 \,  {\rm GeV}\, 
\end{eqnarray}
We will need the meson masses \cite{pdg} 
\begin{equation}
\label{MB}
M_{B_d}=5.279 \, {\rm GeV} \, , 
\qquad 
M_{B_s}=5.368  \, {\rm GeV} \, . 
\end{equation}
The average width $\Gamma_s$ of the $B_s$ mass eigenstates 
is computed from the well-measured $B_d$ lifetime, 
\begin{equation}
\tau_{B_d}  = 1.530 \pm 0.009 \, \mbox{ps} \, ,
\end{equation}
using $\Gamma_s = 1/\tau_{B_d} \, (1.00 \pm 0.01)$. 
Our input of the CKM elements is \cite{exput} 
\begin{eqnarray}
|V_{us}|   & =&   0.2248 \pm 0.0016\, , \qquad
|V_{cb}|   \; =\;  \left( 41.5 \pm 1.0 \right) \cdot 10^{-3}\, \nn 
\left|\frac{V_{ub}}{V_{cb}}\right| & = & 0.10 \pm 0.02\, , 
\, \qquad\qquad\quad
\gamma \; =\;  1.05^{+0.31}_{-0.12} \, .\label{vubnum}
\end{eqnarray}
For all predictions within the standard model we assume unitarity of the
CKM matrix and we determine all CKM elements from the four parameters
$|V_{us}|$, $|V_{cb}|$, $|V_{ub}/V_{cb}|$ and $\gamma$.  The W mass
\cite{pdg} and the strong coupling constant are \cite{b}
\begin{equation}
\label{mw}
M_W=80.4 \, {\rm GeV} \, , \qquad\qquad \alpha_s (M_Z) = 0.1189 \pm 0.0010 
\, .
\end{equation}
We note that in the $B_s$ system CKM parameters other than $|V_{cb}|$
(which basically determines $|V_{ts}|$) play a minor role. The same is
true for the strange quark mass in \eq{ms}.  

The dominant theoretical uncertainties, however, stem from the
non-perturbative parameters discussed below and from the dependence on
the unphysical renormalisation scale $\mu_1$.  We use the central values
$\mu_1 = \mu_2 = m_b$ and we vary $\mu_1$ between $m_b/2$ and $2 m_b$.
The dependence on $\mu_2$ is related to the determination of the
hadronic quantities and uncertainties associated with $\mu_2$ are
contained in the quoted ranges for these quantities.
\\
The situation of the non-perturbative parameters - the decay constant
and the bag parameters - is not yet settled. Different non-perturbative
methods result in quite different numerical results.  QCD sum rule
estimates were obtained for the decay constant $f_{B_s}$
\cite{fBsumrule}, for the bag parameter $B$ \cite{Bsumrule, BSsumrule}
and for $B_S$\cite{BSsumrule}. The same quantities have been determined
in quenched approximation in numerous lattice simulations, see
\cite{latticereview} for a review.  The only determination of
$\tilde{B}_S$ was done in a quenched lattice simulation in \cite{bgmpr}.
Unquenched ($n_f = 2$) values are available for $f_{B_s}$
\cite{fBslatticeun2,Blatticeun}, for $B$ \cite{Blatticeun,Blatticeun2}
and for $B_S$ \cite{Blatticeun2, BSlatticeun}.  For the decay constant
$f_{B_s}$ even a lattice simulation with 2+1 dynamical fermions is
available \cite{fBslatticeun2+1}.
\\
Unfortunately it turns out that the predictions for $f_{B_s}$ vary over
a wide range, ${\cal O} (200 \pm 20 \, \mbox{MeV})$ for quenched
results, ${\cal O} (230 \pm 20 \, \mbox{MeV})$ for $n_f = 2$, ${\cal O}
(245 \pm 20 \, \mbox{MeV})$ for sum rule estimates and ${\cal O} (260
\pm 29 \, \mbox{MeV})$ for $n_f = 2+1$, see e.g.  \cite{latticereview}.
This discrepancy has to be resolved, since $\Delta M$ and $\Delta
\Gamma_s$ depend quadratically on the decay constant!  Recently the
combinations $f_{B_s}^2 B$, $f_{B_s}^2 B_S$ and $f_{B_s}^2 \tilde B_S$
were determined for 2+1 light flavors \cite{latticeall}.  The authors of
\cite{latticeall} claim that the combined determination results in a
considerable reduction of the theoretical error.
\\
We will use in our numerics two sets of non-perturbative parameters:
\\
{\bf Set I} consists of a conservative estimate for $f_{B_s}$ combined
with the unquenched determination for $B $ \cite{Blatticeun} and $B_S$
\cite{BSlatticeun} and the only published lattice determination of
$\tilde B_S$ \cite{bgmpr}:
\begin{eqnarray}
f_{B_s}    & = & 240 \pm 40 \, \mbox{MeV} 
\nn
B    & = & 0.85 \pm 0.06  \quad   \Rightarrow \quad  
                 f_{B_s} \sqrt{B}    = 0.221 (46) \, \mbox{GeV}
\nn
B_S        & = & 0.86 \pm 0.08  \quad \Rightarrow \quad     
B_S^\prime \; =\; 1.34 \pm 0.12 \quad 
\Rightarrow  \quad 
f_{B_s} \sqrt{B_S^\prime} \; =\; 0.277 (57) \, \mbox{GeV}
\nn
\tilde B_S & = & 0.91 \pm 0.08  \quad \Rightarrow \quad 
\tilde B_S^\prime \; =\; 1.41 \pm 0.12  
           \quad \Rightarrow \quad 
              f_{B_s} \sqrt{\tilde B_S'} = 0.285 (60) \, \mbox{GeV}
\label{setone}
\end{eqnarray}
{\bf Set II} consists of the preliminary determination with 2+1 flavors
\cite{latticeall}:
\begin{eqnarray}
f_{B_s} \sqrt{B}        & = & 0.227 (17) \, \mbox{GeV}
\nn
f_{B_s} \sqrt{B_S'}           & = & 0.295 (22) \, \mbox{GeV}
\nn
f_{B_s} \sqrt{\tilde B_S'}    & = & 0.305 (23) \, \mbox{GeV}
\end{eqnarray}
The central values of both sets are quite similar, while the errors of
set II are smaller by almost a factor 3.
\\
For both sets the bag parameters of the $1/m_b$-corrections 
are estimated within vacuum insertion approximation and we use the following 
conservative error estimate
\begin{equation}
B_{R_i} = 1 \pm 0.5 \, . \label{br}
\end{equation}
In our computer programs we carefully extract all terms of order
$\alpha_s^2$ and $\alpha_s/m_b$, which belong to yet uncalculated orders
of the perturbation series, and discard them consistently.
\subsection{$\mathbf{\dm_s}$ within the SM}
In the standard model expression (Eq.(\ref{dmdg}) \& Eq.(\ref{m12sm})) 
for the mass difference in the $B_s$-system a product of perturbative 
corrections ($\hat \eta_B S_0$) and  non-perturbative 
corrections ($ f_{B_s}^2 B$) arises.
Using the above input the perturbative corrections are given by \cite{bjw} 
\begin{eqnarray}
\hat \eta_B(\mu = \ov m_b) & = & 0.837 \, (\mbox{NDR}), 
 \\
S_0(x_t)=S_0\left(\frac{\ov m_t^2(\ov m_t)}{M_W^2}\right) & = &
\frac{4x_t - 11 x_t^2 + x_t^3}{4 (1-x_t)^2} - 
\frac{3 x_t^3 \ln (x_t)}{2 (1-x_t)^3}
=
2.327 \pm 0.044 \label{defs0}
\end{eqnarray}
Our final values for the standard model prediction 
\begin{eqnarray}
\Delta M_s & = & \left( 19.30 \pm 6.68 \right) \, \mbox{ps}^{-1} 
\, \, \, \, \mbox{\bf (Set I)}
\\
\Delta M_s & = & \left( 20.31 \pm 3.25 \right) \, \mbox{ps}^{-1}
\, \, \, \, \mbox{\bf (Set II)}
\label{dmsetII}
\end{eqnarray}
are bigger than the experimental result, but consistent within the errors.
Using $f_{B_s}=230\, \mev$ and the bag parameter from set I, 
one exactly reproduces the experimental value of $\Delta M_s$.
\\
The overall error is made up from the following components:
\begin{center}
\begin{math}
\begin{array}{|c||c|c|}
\hline
\mbox{Input} & \Delta M_s & \Delta M_s
\\
& \mbox{Set I}  & \mbox{Set II}
\\
\hline
\hline
f_{B_s}      & 1^{+0.361}_{-0.306}    &  - 
\\
\hline
\hline
B          & 1 \pm 0.071         &  - 
\\
\hline
f_{B_s}^2 B & 1 \pm  0.341           & 1 \pm 0.150
\\
\hline
\hline
V_{cb}       & 1^{+0.049}_{-0.048} &  1^{+0.049}_{-0.048}
\\
\hline
\alpha_s (M_Z) & 1 \pm 0.020        & 1 \pm 0.020 
\\
\hline
m_t          & 1 \pm 0.018         & 1 \pm 0.018
\\
\hline
\gamma     & 1^{+0.005}_{-0.015} & 1^{+0.005}_{-0.015}
\\
\hline
|V_{ub}/V_{cb}| & 1 \pm 0.005      & 1 \pm 0.005 
\\
\hline
\hline
\sqrt{\sum \bar \delta^2} & 1 \pm 0.346  & 1 \pm 0.160
\\
\hline
\end{array}
\end{math}
\end{center}
When combining different errors we first symmetrised the individual errors and
added them quadratically afterwards.  The by far dominant contribution to the
error comes from the non-perturbative parameter $f_{B_s}^2 B$. Clearly, in
view of the precise measurement in \eq{dmexp} it is highly desirable to
understand the hadronic QCD effects with a much higher precision than today.
\boldmath
\subsection{$\dg_s$, $\dg_s/\dm_s $ and $a_{\rm fs}^s$ within 
            the SM}\label{sect:sm} 
\unboldmath
The main result of this paper is a new, more precise determination of 
$\Gamma_{12}$, which is then used to determine $\Delta \Gamma_s$, 
$\Delta \Gamma_s/ \Delta M_s$ and $a_{\rm fs}^s$.

In order to illustrate our progress, we first present the results in the
old operator basis used in \cite{bbgln1,rome03}.  Using the scheme
involving $m_b^{\rm pole}$ and $z$ as in \cite{bbgln1,rome03}, but
updating the input parameters to our values in \eqsto{mb}{mw}, we find
\begin{eqnarray}
\Delta \Gamma_{s,old}^{\rm pole} & = & 
\left( \frac{f_{B_s}}{240 \, \mbox{MeV}} \right)^2
\left[ 0.002 B + 0.094  B_S' - \rt. \nn
&& \qquad\quad \lt. 
\left(0.033  B_{\tilde{R}_2} +  0.019 B_{R_0}+  0.005 B_R \right) \right]
\, \mbox{ps}^{-1}
\nonumber
\\
\Delta \Gamma_{s,old}^{\rm pole, LO} & = & 
\left( \frac{f_{B_s}}{240 \, \mbox{MeV}} \right)^2
\left[ 0.005 B + 0.145 B_S' - 
\rt. \nn && \qquad\quad \lt.
\left(0.033  B_{\tilde{R}_2} +  0.019 B_{R_0}+  0.005 B_R \right) \right]
\, \mbox{ps}^{-1} 
\nonumber
\\
a_{\rm fs, old}^{\rm pole,s} & = & 
\left[ 10.8  + 1.9 \frac{ B_S'}{B} + 0.8 \frac{B_R}{B}   \right]
\mbox{Im} \left( \frac{\lambda_u}{\lambda_t} \right) \cdot 10^{-4}
\nonumber
\\
& + & 
\left[ 0.10  - 0.01 \frac{ B_S'}{B} + 0.29 \frac{B_R}{B}  \right]
\mbox{Im} \left( \frac{\lambda_u}{\lambda_t} \right)^2 \cdot 10^{-4}
\nn
\left( \frac{\Delta \Gamma_s}{\Delta M_s} \right)_{old}^{\rm pole} & = & 
\left[ 0.9  + 40.9  \frac{ B_S'}{B} 
- \left(14.4 \frac{B_{\tilde{R}_2}}{B} + 8.5 \frac{B_{R_0}}{B} 
+ 2.1 \frac{B_R}{B}  \right) \right] \cdot 10^{-4}
\label{oldr}
\end{eqnarray}
For simplicity we do not show the uncertainties of the numerical coefficients
appearing in the square brackets here and in following similar occasions. We 
assess these uncertainties, however, when quoting final results.   

Several comments are in order: in the old basis the coefficient of $B$ in the
prediction of $\Delta \Gamma_s$ is negligible due to a cancellation among
$\Delta B=1$ Wilson coefficients, thus the term with $B_S^\prime$ dominates
the overall result.  This leads to the undesirable fact that the only
coefficient in $\Delta \Gamma_s / \Delta M_s$ that is free from
non-perturbative uncertainties is numerically negligible.  Moreover in $\Delta
\Gamma_s$ all $1/m_b$-corrections have the same size and add up to an
unexpectedly large correction (30$\%$ of the LO value, 45$\%$ of the
NLO value).  In \eq{oldr} we have singled out the bag factors of the two most
important sub-dominant operators $\tilde R_2$ and $R_0$, while the bag
parameters of the remaining operators are chosen equal and are denoted by
$B_R$.  Finally in the old operator basis the calculated NLO QCD corrections
are large and reduce the final number by about 35$\%$ of the LO value.
\\
$a_{\rm fs}^s$ does not suffer from this shortcomings.  Here the coefficient
without non-perturbative uncertainties is numerically dominant and the size of
the $1/m_b$ corrections seems to be reasonable.  Moreover in this case $R_3$
and $\tilde R_3$ are the dominant subleading operators. Since the overall
contribution of the $1/m_b$-corrections is relatively small, we choose all
bag factors of power suppressed operators equal to $B_R$.
\\
Using the non-perturbative parameters from set I we obtain the following
number for $\Delta \Gamma_s$:
\begin{eqnarray}
\Delta \Gamma_s & = & \left( 0.070   \pm 0.042 \right) \, \mbox{ps}^{-1}
 \hspace{0.25cm} \Rightarrow \hspace{0.25cm}
\frac{\Delta \Gamma_s}{\Gamma_s}  = 
\Delta \Gamma_s \cdot \tau_{B_d} = 0.107 \pm 0.065
\label{dgold}
\end{eqnarray}
This number is in agreement with previous estimates 
\cite{bbgln1,previous1,previous2,previous3}
where different input parameters - in particular different values for the
decay constant and the bag parameters - were used.  In the following table we
quote the central values of these old predictions and in addition give
the corresponding 
results adjusted to the new non-perturbative parameters of set I: 
\begin{center}
\begin{math}
\begin{array}{|c|c|c|c|c|}
\hline
\mbox{Reference} & \mbox{predicted } \Delta \Gamma_s / \Gamma_s 
& \mbox{ used } f_{B_s} & \mbox{ used } B_S' &
\Delta \Gamma_s / \Gamma_s (f_{B_s} = 240 \, \mbox{MeV}, 
\\ 
      &  &   &   & \, \, \, \, \, \, \, \, B_S' = 1.34)
\\
\hline
\mbox{\cite{bbgln1}}        & 0.054 & 210 & 1.02 & 0.117
\\
\hline
\mbox{\cite{previous1}} & 0.093 & 230 & 1.25 & 0.114
\\
\hline
\mbox{\cite{previous2}}      & 0.124 & 245 & 1.36 & 0.116
\\
\hline
\mbox{\cite{previous3}}         & 0.118 & 245 & 1.31 & 0.117
\\
\hline
\end{array}
\label{oldnumbers}
\end{math}
\end{center}
The values in the last column are still bigger than the new number in
Eq.(\ref{dgold}) by about $8\%$.  Besides some differences from other
input parameters --- like quark masses and CKM parameters --- this small
overestimate in the last column originates from the use of different
methods to determine $\Gamma_s$ in the ratio $\Delta \Gamma_s /
\Gamma_s$ compared to this work.  Since now very precise values of the
b-lifetimes are available, we directly use them as an input to determine
the total decay rate: $\Gamma_s = 1/ \tau_{B_d}$. In
\cite{bbgln1,previous1,previous2,previous3} we expressed the total decay
rate in terms of the semileptonic decay rate: $ \Gamma_s = \Gamma_{\rm
  sl}^{\rm theory} / B_{\rm sl}^{\rm exp}$.  Doing so (with the 1998
value of $B_{\rm sl}^{\rm exp}$) one obtains values for $\tau_B \approx
1.66 \, \mbox{ps}$, which are about $8\%$ larger than the experimental
number of $\tau_{B_d} \approx 1.53 \, \mbox{ps}$.
\\
The Rome group \cite{rome03} used a different normalisation, guided by
the wish to eliminate the huge uncertainty due to $f_{B_s}$: $ \Delta
\Gamma_s / \Gamma_s = (\Delta \Gamma_s / \Delta M_s)^{ \rm theory}
(\Delta M_s /\Delta M_d)^{\rm theory} \Delta M_d^{\rm exp} \tau_{B_s}$.
The values obtained by the Rome group for $\Delta \Gamma_s/ \Gamma_s$
were typically considerably lower than 0.10, which was partially due to
different input parameters like the bottom mass.  Since now $\Delta M_s$
is known experimentally one can abbreviate their method to $ \Delta
\Gamma_s / \Gamma_s = (\Delta \Gamma_s / \Delta M_s)^{\rm theory} \Delta
M_s^{\rm exp} \tau_{B_s}$.  This prediction assumes that no new physics
effects contribute to the mass difference.  This is numerically
equivalent to the use of $f_{B_s}=230\; \mev$ in our approach (see the
passage below Eq. (\ref{dmsetII})).  With that input we obtain from our
analysis $\Delta \Gamma_s / \Gamma_s = 0.10 \pm 0.06$ which is in
perfect agreement with the latest update of the Rome group from this
year \cite{rome06}.  Thus we see no discrepancy anymore between our
predictions and those of the Rome group.

However, our predictions have been criticised recently in
\cite{dmphenburas}.  The authors of \cite{dmphenburas} obtain a much lower
central value - $\Delta \Gamma_s / \Gamma_s = 0.067 \pm 0.027$ - and
claim that this difference stems from their use of lattice values for
the $1/m_b$-operators, while in our approach the vacuum insertion
approximation was used. Lattice values for the $1/m_b$ corrections can
be extracted from \cite{bgmpr} for the operators $R_0, R_1$ and $\tilde
R_1$, but their use does not resolve the numerical discrepancy.
With the help of one author of \cite{dmphenburas} we have traced the
difference back to the omission of the radiative corrections contained
in $\alpha_1$ and $\alpha_2$, when \eq{brn} is used to extract $\langle
R_0 \rangle$ from lattice data on $\langle Q\rangle$, $\langle
Q_S\rangle$ and $\langle \widetilde{Q}_S \rangle$.  This is numerically
equivalent to shifting $B_{R_0}$ from 1.1 to 1.7.  If we use this number
and $f_{B_s}=230\; \mev$ we obtain $\Delta \Gamma_s / \Gamma_s = 0.079$,
which is closer to but still larger by $18\%$ than the value obtained in
\cite{dmphenburas}.


Now we turn to the results in the new basis: For a direct comparison
with the old operator basis, we first show results for the scheme 
characterised by $m_b^{\rm pole}$ and $z$:
\begin{eqnarray}
\Delta \Gamma_s^{\rm pole} & = & 
\left( \frac{f_{B_s}}{240 \, \mbox{MeV}} \right)^2
\left[ 0.095 B + 0.023 \tilde B_S' -
\rt. \nn && \qquad\quad \lt. 
\left(0.033 B_{\tilde{R}_2} - 0.006 B_{R_0}+ 0.005 B_R \right) \right]
\, \mbox{ps}^{-1}
\nonumber
\\
\Delta \Gamma_s^{\rm pole, LO} & = & 
\left( \frac{f_{B_s}}{240 \, \mbox{MeV}} \right)^2
\left[  0.121 B + 0.029 \tilde B_S' - 
\rt. \nn && \qquad\quad \lt.
\left(0.033 B_{\tilde{R}_2} - 0.006 B_{R_0}+ 0.005 B_R \right) \right]
\, \mbox{ps}^{-1} \nonumber
\\
a_{\rm fs}^{\rm pole, s} & = & 
\left[  12.9 +  0.5 \frac{\tilde B_S'}{B} +  1.7 \frac{B_R}{B}  \right]
\mbox{Im} \left( \frac{\lambda_u}{\lambda_t} \right) \cdot 10^{-4}
\nonumber
\\
& + & 
\left[ 0.20  +  0.02 \frac{\tilde B_S'}{B} + 0.44 \frac{B_R}{B} \right]
\mbox{Im} \left( \frac{\lambda_u}{\lambda_t} \right)^2 \cdot 10^{-4}
\\
\left( \frac{\Delta \Gamma_s}{\Delta M_s} \right)^{\rm pole} & = & 
\left[ 41.4 +  10.0  \frac{\tilde B_S'}{B} 
- \left(14.4 \frac{B_{\tilde{R}_2}}{B} -  2.6 \frac{B_{R_0}}{B} + 
2.1 \frac{B_R}{B}  \right) \right] \cdot 10^{-4}
\end{eqnarray}
Now we are in the desired situation that $\Delta \Gamma_s$ is dominated
by $B$ and the lion's share of $\Delta \Gamma_s / \Delta M_s$ can be
determined without any hadronic uncertainty!  Moreover the size of the
$1/m_b$-corrections has become smaller, because the magnitude of the
contribution from $R_0$ is reduced by a factor of 3 (as anticipated from
\eqsand{gcc}{guc}) and the sign of this contribution has changed. We are
left with a $1/m_b$ correction of 22$\%$ of the LO value or 28$\%$ of
the NLO-value.  Using the new operators the $\alpha_s$-corrections have
become smaller (22$\%$ of the LO value), too, and the unphysical
$\mu_1$-dependence has shrunk.  In the case of $a_{\rm fs}^s=\imag
(\Gamma_{12}^s/M_{12}^s)$ the situation did not change much due to the
change of the basis. Here we have no strong recommendation on what basis
to choose. However, in the presence of new physics $a_{\rm fs}^s$ also
involves $\real (\Gamma_{12}^s/M_{12}^s)$ and the same improvements
occur, as discussed in Sect.~\ref{sect:new}.
\\
Using the non-perturbative parameters from set I we obtain the following
number for $\Delta \Gamma_s$:
\begin{eqnarray}
\Delta \Gamma_s & = & \left( 0.081   \pm 0.036 \right) \, \mbox{ps}^{-1}
 \hspace{0.25cm} \Rightarrow \hspace{0.25cm}
\frac{\Delta \Gamma_s}{\Gamma_s}  = \Delta \Gamma_s \cdot \tau_{B_d} 
= 0.124 \pm 0.056
\label{dgnewpz}
\end{eqnarray}
The central value in the new basis is larger than the old one, while the
theoretical errors have shrunk considerably.  The numerical difference stems
from uncalculated corrections of order $\alpha_s/m_b$ and $\alpha_s^2$.  As
a consistency check of our change of basis one can compare the results in the
old and the new basis neglecting all $1/m_b$ and $\alpha_s$-corrections and
setting $B=1=B_S'$.  As required we get in both cases the same result: $\Delta
\Gamma_s / \Gamma_s = 0.1497$.

For our final number we still go further. First we sum up logarithms of the
form $z \ln z$ by switching to schemes using $\ov z$ defined in \eq{defzb}.
Second we calculate our results for two schemes of the b-quark mass, using
either $\ov m_b$ or $m_b^{\rm pole}$ of \eq{mb} and finally 
average over the schemes. By this we obtain the main result of this paper:
\begin{eqnarray}
\Delta \Gamma_s & = & \left( \frac{f_{B_s}}{240 \, \mbox{MeV}} \right)^2
\left[ (0.105 \pm 0.016) B + (0.024 \pm 0.004)  \tilde B_S' \right. 
\nonumber 
\\ 
&& 
\left. \hspace{1.2cm} - 
\left( (0.030 \pm 0.004) B_{\tilde{R}_2} - 
       (0.006\pm 0.001) B_{R_0} + 
        0.003 B_R \right) \right] \, \mbox{ps}^{-1} \quad
\label{finaldg}
\\
a_{\rm fs}^s & = & 
\left[ (9.7 \pm 1.6) + 0.3 \frac{\tilde B_S'}{B} + 0.3 \frac{B_R}{B}  \right]
\mbox{Im} \left( \frac{\lambda_u}{\lambda_t} \right) \cdot 10^{-4}
\nonumber
\\
& + & 
\left[ (0.08 \pm 0.01)  
      + 0.02 \frac{\tilde B_S'}{B} + (0.05 \pm 0.01) \frac{B_R}{B} \right]
\mbox{Im} \left( \frac{\lambda_u}{\lambda_t} \right)^2 \cdot 10^{-4}
\\
\frac{\Delta \Gamma_s}{\Delta M_s}  & = & 
\left[ (46.2 \pm 4.4)   + (10.6 \pm 1.0) \frac{\tilde B_S'}{B} \right.
\nonumber \\ && \left.
- \left( (13.2 \pm 1.3) \frac{B_{\tilde{R}_2}}{B} 
        - (2.5 \pm 0.2) \frac{B_{R_0}}{B} 
        + (1.2 \pm 0.1) \frac{B_R}{B}  \right) \right] \cdot 10^{-4}
\end{eqnarray}
Using the parameter set I, we obtain the following final numbers
\begin{eqnarray}
\Delta \Gamma_s & = & \left( 0.096   \pm 0.039 \right) \, \mbox{ps}^{-1}
 \hspace{0.25cm} \Rightarrow \hspace{0.25cm}
\frac{\Delta \Gamma_s}{\Gamma_s}  = \Delta \Gamma_s \cdot
 \tau_{B_d} = 0.147 \pm 0.060 \label{findg}
\\
a_{\rm fs}^s & = & \left( 2.06 \pm 0.57 \right) \cdot 10^{-5}
\\
\frac{\Delta \Gamma_s}{\Delta M_s}  & = & 
\left( 49.7 \pm 9.4 \right) \cdot 10^{-4}
\\
\phi_s & = & (4.2\pm 1.4 )\cdot 10^{-3}
      \; = \; 0.24^\circ \pm 0.08^\circ 
\label{finphi}
\end{eqnarray}
The first striking feature of these numbers is the large increase for
the prediction of $\Delta \Gamma_s$ from 0.070 $\mbox{ps}^{-1}$ to 0.096
$\mbox{ps}^{-1}$ (about 37 $\%$). The change of the basis is responsible
for an increase of about 16 $\%$. We have shown that the previously used
basis suffers from several serious drawbacks --- most importantly in the
old basis strong cancellations, which are absent in the new basis,
occur.  Next we have reduced an additional uncertainty by summing up
logarithms of the form $z \ln z$ to all orders. This theoretical
improvement results in another increase of about $11 \%$.  The averaging
over the pole and $\overline{\mbox{MS}}$ schemes results in an increase
of about $7 \%$ compared to the exclusive use of the pole-scheme.
Finally we also include subleading CKM-structures (as done in
\cite{rome03,bbln} as well) giving an increase of $\Delta \Gamma_s$ by
about $3\%$ compared to setting $V_{ub}$ to zero. In the case of the
flavour-specific CP-asymmetry the choice of the new basis has no
dramatic effect.
\\
If one assumes that there is no new physics in the measured value of
$\Delta M_s$ one can avoid the large uncertainty due to $f_{B_s}$ by
writing:
\begin{eqnarray}
\Delta \Gamma_s & = & 
\left( \frac{\Delta \Gamma_s}{\Delta M_s} \right)^{\rm Theory }
\cdot \Delta M_s^{\rm Exp.} = 0.088 \pm 0.017  
\, \mbox{ps}^{-1} 
\\
& \Rightarrow &
\frac{\Delta \Gamma_s}{\Gamma_s} 
= \Delta \Gamma_s \cdot \tau_{B_d} = 0.127 \pm 0.024 \, .
\end{eqnarray}
This smaller value is numerically equivalent to using $f_{B_s} = 230$ MeV in 
\eq{finaldg}.
\\
For completeness we also present the numbers with the parameter set II:
\begin{eqnarray}
\Delta \Gamma_s & = & \left( 0.106  \pm 0.032 \right) \, \mbox{ps}^{-1}
\hspace{0.25cm} \Rightarrow \hspace{0.25cm}
\frac{\Delta \Gamma_s}{\Gamma_s} =
\Delta \Gamma_s \cdot \tau_{B_d} = 0.162 \pm 0.049
\\
a_{\rm fs}^s & = & \left( 2.06 \pm 0.57 \right) \cdot 10^{-5}
\\
\frac{\Delta \Gamma_s}{\Delta M_s}  & = & 
\left(51.9\pm 9.8 \right) \cdot 10^{-4}
\end{eqnarray}
The above errors in $\Delta \Gamma_s$ and $\Delta M_s$ have to be taken with
some care, since we were not using our conservative error estimate but the
preliminary values from \cite{latticeall}.

In the following table the individual sources of uncertainties in $\Delta
\Gamma_s$ --- using the parameter set I --- are listed in detail:
\begin{center}
\begin{equation}
\begin{array}{|c||c|c|c||}
\hline
\mbox{Input} & \Delta \Gamma_s & \Delta \Gamma_s & \Delta \Gamma_s 
\\
 & \mbox{old, pole, z} & \mbox{new, pole, z} & \mbox{new, average, } \ov z
\\
\hline
\hline
f_{B_s}      & 1^{+0.361}_{-0.306} & 1^{+0.361}_{-0.306} &1^{+0.361}_{-0.306}
\\
\hline
\hline
B_1          & 1 \pm 0.002         &  1^{+0.070}_{-0.071}& 1 \pm 0.066 
\\
\hline
B_{2,3}      & 1 \pm 0.167         &  1 \pm 0.035& 1 \pm 0.031
\\
\hline
B_{\tilde{R}_2}  & 1 \pm 0.235     &  1 \pm 0.203& 1 \pm 0.157
\\
\hline
B_{R_0}      & 1 \pm 0.140         &  1 \pm 0.036& 1 \pm 0.030
\\
\hline
\mu_1 \mbox{ with }m_b/2 \leq \mu_1 \leq 2m_b        
      & 1^{+0.248}_{-0.521} & 1^{+0.111}_{-0.272}& 1^{+0.074}_{-0.200}
\\
\hline
V_{cb}       & 1^{+0.049}_{-0.048} & 1^{+0.049}_{-0.048}& 1 \pm 0.049
\\
\hline
z            & 1^{+0.044}_{-0.046} & 1^{+0.040}_{-0.042}& 1 \pm 0.019
\\
\hline
m_b          & 1^{+0.043}_{-0.042} & 1^{+0.036}_{-0.035}& 1^{+0.010}_{-0.009}
\\
\hline
\alpha_s     & 1^{+0.014}_{-0.013} & 1 \pm 0.003& 1 \pm 0.001
\\
\hline
m_s          & 1 \pm 0.010         & 1 \pm 0.012 & 1 \pm 0.010
\\
\hline
\gamma       & 1^{+0.005}_{-0.016} & 1^{+0.005}_{-0.015}& 1^{+0.005}_{-0.014}
\\
\hline
|V_{ub}/V_{cb}| & 1 \pm 0.006      & 1 \pm 0.006 & 1 \pm 0.005
\\
\hline
\hline
\sqrt{\sum \bar \delta^2} & 1 \pm 0.607  & 1 \pm 0.450 & 1 \pm 0.405 
\\
\hline
\hline
m_b^{\rm pow}  & 1_{-0.368}          & 1_{-0.158}&  1_{-0.112}
\\
\hline
\mbox{RS} & 1 \pm 0.133 & 1 \pm 0.065 & 1 \pm 0.066 
\\
\hline
\end{array}
\label{tabledg}
\end{equation}
\end{center}
The same result is visualised in figure \ref{kuchendg}.
\begin{nfigure}{tb}
\includegraphics[width=0.9\textwidth,angle=0]{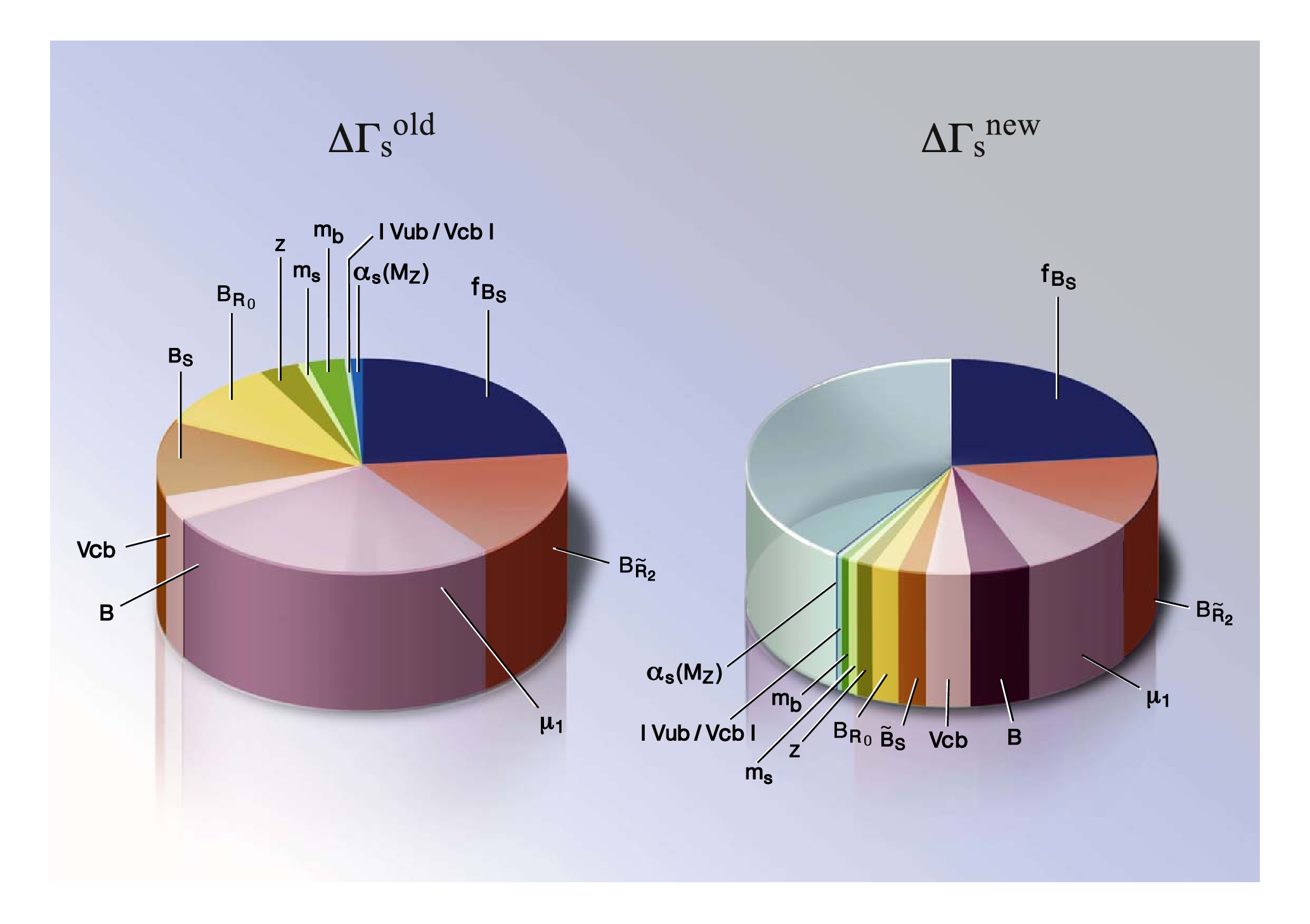}
\caption{Uncertainty budget for the theory prediction of $\dg_s$. 
The largest uncertainties stem from $f_{B_s}$, the renormalisation scale
  $\mu_1$ of the $\Delta B=1$ operators and the bag parameter of the 
  $1/m_b$--suppressed operator $\widetilde{R}_2$. The transparent 
  segment of the right pie chart shows the improvement with respect to 
  the old result on the left.}\label{kuchendg}
\end{nfigure}
In the case of $\Delta \Gamma_s$ the by far largest uncertainty stems from the
error on $f_{B_s}$. Here a considerable improvement from the non-perturbative
side is mandatory. The dependence on the decay constant is of course not
affected by the change of the operator basis.  The second most important
uncertainty comes from the $1/m_b$-operator $\tilde{R}_2$.  This operator has
up to now only been estimated in the naive vacuum insertion approximation. Any
non-perturbative investigation would be very helpful.
Number three in the error hit list is the unphysical $\mu_1$-dependence. Using
the old operator basis the corresponding error was huge, it was drastically
reduced by changing to the new basis and by including also the
$\overline{\mbox{MS}}$-scheme for the b-quark mass. Any further improvement
requires a cumbersome NNLO calculation, which might be worthwhile if progress
on the non-perturbative side for $f_{B_s}$ and $\tilde{R}_2$ is achieved.
Number four is again a non-perturbative parameter - now the bag parameter of
the operator $Q$. In the old operator basis the corresponding uncertainty
stemmed from $B_S$ and was larger by a factor of 2.5.  The dependence on
$V_{cb}$ results in a relative error of about $5\%$ for both the old basis and
the new basis. All remaining uncertainties are at most $3\%$.
\\
Using our conservative estimates and adding all errors quadratically (after
symmetrising them) we arrive at a reduction of the overall theoretical error
due to the introduction of the new basis from $\pm 61\%$ to $ \pm 41 \%$,
where the last number is completely dominated by the decay constant.  If one
neglects the dependence on $f_{B_s}$ the overall theoretical error goes down
from $\pm 51\%$ to $ \pm 23 \%$.
\\
In the table in \eq{tabledg} we also show the dependence on the b-quark mass we
are using in the $1/m_b$-corrections, $m_b^{\rm pow}$. This dependence can be
viewed as a measure of the overall size of the $1/m_b$-corrections. The use of
the new basis results in a strong reduction of the corresponding uncertainty,
from $37\%$ to $11\%$.  And finally we compare the two renormalisation
schemes (RS) we are using for the b-quark mass. Here we have again muss less
uncertainty in the new operator basis.  To avoid a double counting of the
errors we did not include the last two rows of table (\ref{tabledg}) in the
total error.
\\
Investigating the case of $\Delta \Gamma_s / \Delta M_s$ the improvement due
to our new basis is more substantial, since here the dependence on
$f_{B_s}$ cancels:
\begin{center}
\begin{math}
\begin{array}{|c||c|c||c|}
\hline
\mbox{Input} & \Delta \Gamma_s / \Delta M_s &\Delta \Gamma_s / \Delta M_s 
& a_{\rm fs}^s
\\
& \mbox{old, pole, z}& \mbox{new, average, } \ov z & \mbox{new, average, } \ov z 
\\
\hline
\hline
B_1              & 1^{+0.074}_{-0.064} & 1 \pm 0.005 & 1^{+0.006}_{-0.005}
\\
\hline
B_{2,3}          & 1 \pm 0.167         & 1 \pm 0.031 &  1 \pm 0.004
\\
\hline
B_{\tilde{R}_2}  & 1 \pm 0.235         & 1 \pm 0.157 &  1 \pm 0.025 (\tilde{R}_3)
\\
\hline
B_{R_0}          & 1 \pm 0.140         & 1 \pm 0.030 &   1 \pm 0.011 (R_3)
\\
\hline
\mu_1 \mbox{ with }m_b/2 \leq \mu_1 \leq 2m_b 
& 1^{+0.194}_{-0.495} & 1^{+0.027}_{-0.154} & 1^{+0.152}_{-0.101}
\\
\hline
V_{cb}           & 1 \pm 0.000         & 1\pm 0.000  & 1 \pm 0.000
\\
\hline
z                & 1^{+0.044}_{-0.046} & 1 \pm 0.019& 1^{+0.094}_{-0.092}
\\
\hline
m_b              & 1^{+0.043}_{-0.042} & 1^{+0.010}_{-0.009}& 1^{+0.037}_{-0.036}
\\
\hline
m_t              & 1 \pm 0.018         & 1 \pm 0.018& 1 \pm 0.018
\\
\hline
\alpha_s         & 1 \pm 0.012         & 1 \pm 0.001& 1 \pm 0.007
\\
\hline
m_s              & 1 \pm 0.010         & 1 \pm 0.010 &1 \pm 0.001 
\\
\hline
\gamma           & 1^{+0.001}_{-0.003} & 1^{+0.000}_{-0.001}& 1^{+0.144}_{-0.081}
\\
\hline
|V_{ub}/V_{cb}|  & 1 \pm 0.001         & 1 \pm 0.001& 1^{+0.194}_{-0.196} 
\\
\hline
\hline
\sqrt{\sum \bar \delta^2} & 1 \pm 0.480& 1 \pm 0.189& 1 \pm 0.279 
\\
\hline
\hline
m_b^{\rm pow}      &  1_{-0.368}         &1_{-0.112}& 1^{+0.016}
\\
\hline
\mbox{RS}        & 1 \pm 0.136 & 1 \pm 0.069& 1 \pm 0.004 
\\
\hline
\end{array}
\end{math}
\end{center}

\begin{nfigure}{tb}
\includegraphics[width=0.9\textwidth,angle=0]{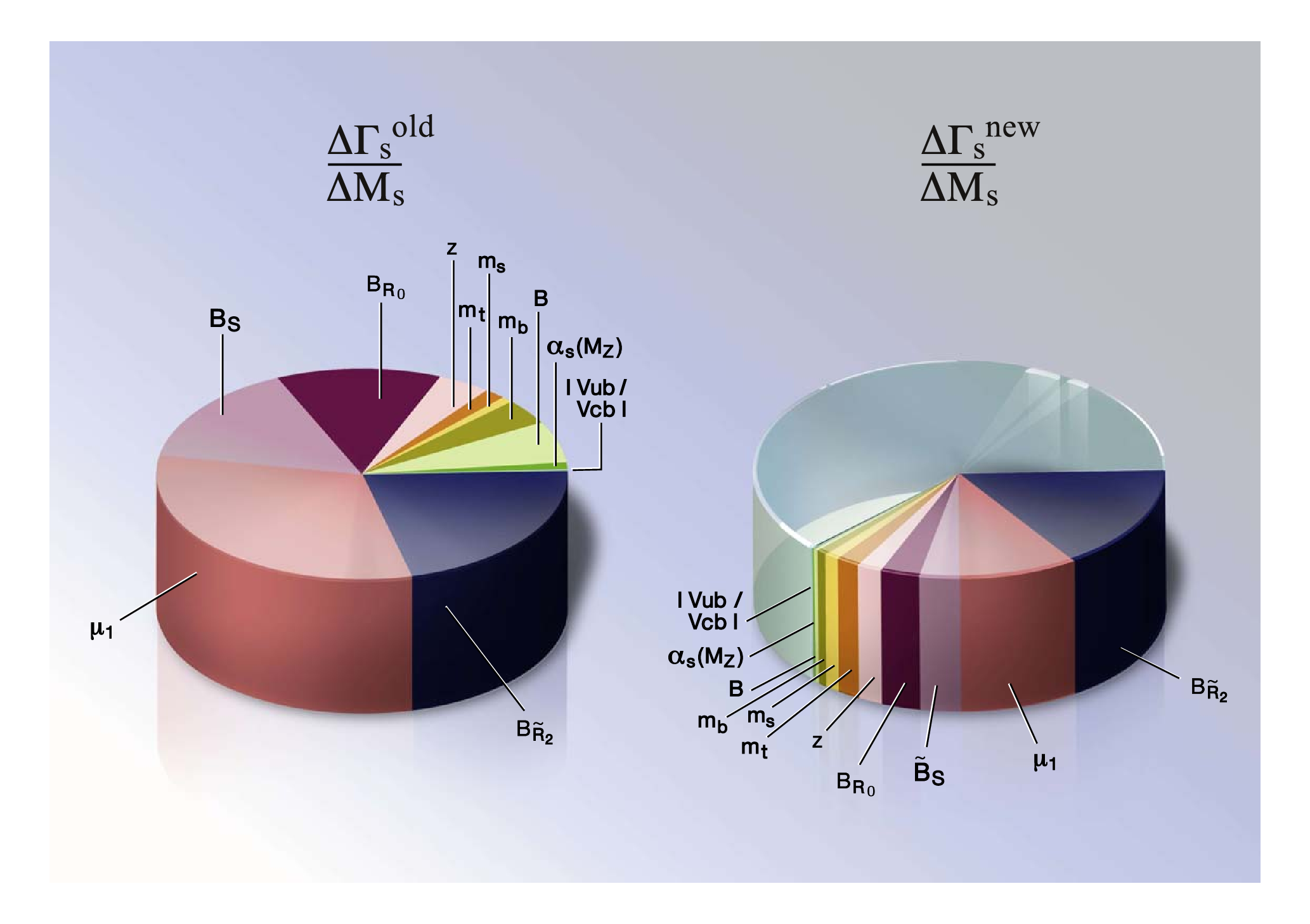}
\caption{Uncertainty budget for $\dg_s/\dm_s$. 
  See \fig{kuchendg} for explanations. 
  The ratio $\dg_s/\dm_s$ does not depend on $f_{B_s}$ and the progress 
  due to the new operator basis is more substantial than in 
  $\dg_s$.}\label{Kuchendgdm} 
\end{nfigure}

In the case of $\Delta \Gamma_s / \Delta M_s$ the use of the new
operator basis leads to a reduction of the total error from $48\%$ to
$19\%$!  The dominant error is now due to the bag parameter
$B_{\tilde{R}_2}$, followed by the $\mu_1$-dependence. The remaining
uncertainties are at most $3\%$.  In the case of $a_{\rm fs}^s$ the
situation is quite different. Here the dominant uncertainty stems
from $V_{ub}$, followed by the dependences on $\mu_1$, $\gamma$
and $\ov z$.
Moreover the $1/m_b$-corrections play a minor role here --- as can be read
off from the error due to the variation of $m_b^{\rm pow}$.
\boldmath
\subsection{$\dm_d$, $\dg_d$ and $a_{\rm fs}^d$ within the SM} 
\unboldmath%
Here we give updated numbers for the mixing parameters of the $B_d$
system. The CKM elements governing \bbmd\ appear in the combinations
$\lambda_i^d=V_{id}^* V_{ib}$ for $i=u,c,t$. 
The bag parameters multiplying $f_{B_d}$ below refer to $B_d$
mesons and are different from those in the $B_s$ system.  However, no
non-perturbative computation has shown any numerically relevant
deviation of $B_{B_d}/B_{B_s}$ from 1. 

Updating $\dm_d$ to $\ov m_t (\ov m_t) = 163.8 \pm 2.0\, {\rm GeV}$ gives 
\begin{displaymath}
\dm_d = (0.53\pm 0.02) \, \mbox{ps}^{-1}
          {\lt( \frac{|V_{td}|}{0.0082}\rt)^2 } \;
          \lt( \frac{f_{B_d}}{200 \, \mbox{MeV}} \rt)^2
             \frac{B}{0.85} .
\end{displaymath}
While in the $B_s$ system the values of $\gamma$ and $|V_{ub}|$ in
\eq{vubnum} play a minor role, their uncertainties are an issue for
$\dg_d$ and $a_{\rm fs}^d$. The master formulae are \cite{bbln}  
\begin{eqnarray}
\frac{\dg_d}{\dm_d} &=& - \, 10^{-4} \lt[ c \; +\;  
     a \, \real \frac{\lambda_u^d}{\lambda_t^d} \; +\;  
     b \, \real \frac{\lambda_u^{d\,2}}{\lambda_t^{d\, 2}} 
  \rt] \label{defabc}\\
a_{\rm fs}^d 
     &=& 10^{-4} \lt[ 
       a \, \imag \frac{\lambda_u^d}{\lambda_t^d} \; +\;  
       b \, \imag \frac{\lambda_u^{d\,2}}{\lambda_t^{d\, 2}} 
  \rt] . \label{defab}
\end{eqnarray}
The coefficients  
\begin{eqnarray}
\qquad
a &=& 2\cdot 10^4 \,  \frac{\Gamma_{12}^{uc}-
                 \Gamma_{12}^{cc}}{M_{12}^d/\lambda_t^{d\, 2}},\qquad
b \;=\; 10^4 \, \frac{2 \Gamma_{12}^{uc}-\Gamma_{12}^{cc}
            -\Gamma_{12}^{uu}}{M_{12}^d/\lambda_t^{d\, 2}} 
\nn 
\mbox{and }\quad
c &=& - 10^4 \, \frac{\Gamma_{12}^{cc}}{M_{12}^d/\lambda_t^{d\, 2}}
\label{abc}
\end{eqnarray}
are independent of CKM elements because of
$M_{12}^d\propto\lambda_t^{d\, 2}$. In our new operator basis these 
coefficients read
\begin{eqnarray}
a & = & 9.68 \epm{1.53}{1.48} \; +\;  
        \lt( 0.31 \epm{0.09}{0.07}\rt) 
                \frac{\widetilde{B}_S^\prime}{B} \; + \;
        \lt( 0.27 \epm{0.15}{0.06} \rt) \frac{B_R}{B} \nn
b & =& 0.08 \pm 0.03 \; +\;  
        \lt(0.02 \pm 0.01 \rt) \frac{\widetilde{B}_S^\prime}{B} 
        \; + \; \lt( 0.04 \epm{0.03}{0.01} \rt) \frac{B_R}{B} \nn 
c & = & 
       -46.1 \pm 6.6 \; -\;  
        \lt( 10.5\pm 1.3 \rt) 
                   \frac{\widetilde{B}_S^\prime}{B}  \; +\;  
        \lt( 8.7\epm{4.9}{1.0} \rt)  \frac{B_R}{B} . \no 
\end{eqnarray}
With the hadronic parameters of Set I in \eq{setone} one finds
\begin{eqnarray}
a & = & 10.5 \epm{1.8}{1.7},\qquad 
b\; =\; 0.2 \pm 0.1 , \qquad
c \; =\;  -53.3 \epm{12.7}{11.4} \label{abcnum}
\end{eqnarray}
It is convenient to express $\lambda_u^d/\lambda_t^d$ in
\eqsand{defabc}{defab} in terms of the angle
$\beta=\arg(-\lambda_t^d/\lambda_c^d)$ of the unitarity triangle and the
length $R_t=|\lambda_t^d/\lambda_c^d|$ of the adjacent side \cite{bbln}:
\begin{eqnarray} 
\real \frac{\lambda_u^d}{\lambda_t^d} & = & \frac{\cos\beta}{R_t} - 1, 
\qquad 
\real \frac{\lambda_u^{d\, 2}}{\lambda_t^{d\, 2}} 
 \; = \; \frac{\cos(2\beta)}{R_t^2} - 2 \frac{\cos\beta}{R_t} + 1, \nn
\imag \frac{\lambda_u^d}{\lambda_t^d} & = & - \frac{\sin\beta}{R_t}, 
\qquad 
\imag \frac{\lambda_u^{d\, 2}}{\lambda_t^{d\, 2}} 
 \; = \; - \frac{\sin(2\beta)}{R_t^2} + 2 \frac{\sin\beta}{R_t}. 
\label{lart}
\end{eqnarray}
Clearly the terms involving $\lambda_u^{d\, 2}/\lambda_t^{d\, 2}$ in
\eqsand{defabc}{defab} 
are numerically irrelevant in view of the smallness of $b$.
Moreover, in the preferred region of the Standard Model fit of the
unitarity triangle one has $\cos\beta\approx R_t$, so that $\real
\lambda_u^d/\lambda_t^d$ is suppressed. Setting $a$ and $b$ to zero
in \eq{defabc} reproduces $\dg_d/\dm_d$ within 2\% \cite{bbln} and 
$\dg_d/\dm_d$ is essentially free of CKM uncertainties. 

Inserting \eqsand{abcnum}{lart} into \eqsand{defabc}{defab} yields 
\begin{eqnarray}
\frac{\dg_d}{\dm_d} &=& \lt[ 
   53.3 \epm{11.4}{12.7}
  \; + \;  \lt( 10.3\epm{1.8}{1.7} \rt) \,  
            \lt( 1 \, -\,     \frac{\cos(\beta)}{R_t} \rt) \rt. \nn 
&& \lt.  \qquad \qquad \; + \;  (0.2\pm 0.1)\, 
   \lt( \frac{\cos(\beta)}{R_t} - \frac{\cos(2\beta)}{R_t^2}\rt) \rt] 
    \cdot 10^{-4} 
    \label{dgdmg}
\end{eqnarray}
\begin{eqnarray}
a_{\rm fs}^d &=& - \lt[ \lt( 10.1\epm{1.8}{1.7}\rt) 
               \, \frac{\sin\beta}{R_t} 
               \; +\; (0.2 \pm 0.1 )\, \frac{\sin(2\beta)}{R_t^2} \rt] 
               \cdot 10^{-4} \label{afsg}
\end{eqnarray}
Next we insert the numerical values for $\beta$ and $R_t$ from
\cite{exput}. Since we are interested in testing the hypothesis of new
physics in \bbms, we take values for $\beta$ and $R_t$ obtained prior to
the measurement of $\dm_s$. With $\beta=23^\circ\pm 2^\circ$ and 
$R_t =0.86\pm 0.11$, which correspond to a CL of 2$\sigma$, one finds
\begin{eqnarray}
\frac{\dg_d}{\dm_d} &=& \lt( 52.6 \epm{11.5}{12.8} \rt) 
             \, \cdot 10^{-4}, \qquad\qquad
a_{\rm fs}^d \; = \; \lt(  -4.8\epm{1.0}{1.2} \rt) \, 
                \cdot 10^{-4}. \label{afsnum} 
\end{eqnarray}
Thus these predictions allow for new physics in $\dm_s$, but assume 
that all other quantities entering the standard fit of the unitarity
triangle in \cite{exput} are as in the Standard Model.  
Using $\dm_d^{\rm exp}= 0.507 \pm 0.004 \, {\rm ps}^{-1}$ and 
$\tau_{B_d}^{\rm exp}  = 1.530 \pm 0.009 $ we find from \eq{afsnum}:
\begin{eqnarray}
\dg_d &=& \frac{\dg_d}{\dm_d} \, \dm_d^{\rm exp} \; =\; 
     \lt( 26.7 \epm {5.8}{6.5}\rt) \cdot 10^{-4} \; 
       {\rm ps}^{-1}, \qquad 
\frac{\dg_d}{\Gamma_d} \;=\; \lt( 40.9 \epm{8.9}{9.9} \rt) \cdot 10^{-4}
.   
 \label{dgdnum}
\end{eqnarray}
The result in \eq{dgdnum} is consistent with our prediction in
\cite{bbln}, but the central value is substantially higher. This is not
solely caused by our new operator basis, but also by the use of a
different renormalisation scheme. In both \cite{bbln} and this work we
average over two schemes, but in one of the schemes used in \cite{bbln}
the $z \ln z$ terms are not summed to all orders. Note that the quoted error
of $a_{\rm fs}^d$ in \cite{bbln} corresponds to the 1$\sigma$ ranges of
$\beta$ and $R_t$, while in \eq{afsnum} more conservative 2$\sigma$
intervals have been used. The ranges in \eq{afsnum} imply for the 
CP-violating phase  $\phi_d=\arg(-M_{12}^d/\Gamma_{12}^d)$:
\begin{eqnarray}
\phi_d &=&  -0.091\epm{0.026}{0.038} \; =\; 
            -5.2^\circ\epm{1.5^\circ}{2.1^\circ} \, . 
\end{eqnarray}

\boldmath
\section{Constraining new physics with \bbms}
\label{sect:new}
\unboldmath
In this section we investigate effects of new physics contributions to
the $B_s$-mixing parameters.
New physics can change the magnitude and the phase of $M_{12}^s$.
We parameterise its effect (similarly to \cite{nir,exput}) by
\begin{eqnarray}
M_{12}^s & \equiv & M_{12}^{\rm SM,s} \cdot  \Delta_s \, ,
\qquad\qquad  \Delta_s \; \equiv \;  |\Delta_s| e^{i \phi^\Delta_s} . 
 \label{defdel}
\end{eqnarray}
The relationship to the parameters used in \cite{nir2006,exput} is
\begin{eqnarray}
\Delta_s &=& r_s^2 e^{2 i \theta_s}. \no
\end{eqnarray}
We find it more transparent to plot $\imag \Delta_s$ vs.\ $\real
\Delta_s$ than to plot $2\theta_s$ vs.\ $r_s^2$. Our plots are similar
to Fig.~1 of \cite{nir2006}, which displays $\sin(2\theta_s)$
vs.~$\cos(2\theta_s)$, but also include the information on
$|\Delta_s|\equiv r_s^2$.  Finally $\Gamma_{12}^s$ stems from
CKM-favoured tree decays and one can safely set $\Gamma_{12}^s =
\Gamma_{12}^{\rm SM,s}$.

\subsection{$\dg_s$, $\dg_s/\dm_s $ and $a_{\rm fs}^s$ beyond
  the SM}\label{sect:by} 
One easily finds: 
\begin{eqnarray}
\dm_s  & = & \dm_s^{\rm SM} \,  |\Delta_s|  
=
(19.30 \pm 6.74 ) \, \mbox{ps}^{-1} \cdot | \Delta_s| 
\label{bounddm}
\\
\Delta \Gamma_s  & = & 2 |\Gamma_{12}^s|
     \, \cos \left( \phi_s^{\rm SM} + \phi^\Delta_s \right)
= (0.096 \pm 0.039) \, \mbox{ps}^{-1} 
\cdot \cos \left( \phi_s^{\rm SM} + \phi^\Delta_s \right)
\label{bounddg}
\\
\frac{\Delta \Gamma_s}{\Delta M_s} 
&= &
 \frac{|\Gamma_{12}^s|}{|M_{12}^{\rm SM,s}|} 
\cdot \frac{\cos \left( \phi_s^{\rm SM} + \phi^\Delta_s \right)}{|\Delta_s|}
=
\left( 4.97 \pm 0.94 \right) \cdot 10^{-3} 
\cdot \frac{\cos \left( \phi_s^{\rm SM} + \phi^\Delta_s \right)}{|\Delta_s|}
\label{bounddgdm}
\\
a_{\rm fs}^s 
&= &
 \frac{|\Gamma_{12}^s|}{|M_{12}^{\rm SM,s}|} 
\cdot \frac{\sin \left( \phi_s^{\rm SM} + \phi^\Delta_s \right)}{|\Delta_s|}
= \left( 4.97 \pm 0.94 \right) \cdot 10^{-3} 
\cdot \frac{\sin \left( \phi_s^{\rm SM} + \phi^\Delta_s
  \right)}{|\Delta_s|} 
\label{boundafs} \\
\lefteqn{\mbox{with (cf.\ \eq{finphi})}
 \qquad \phi_s^{\rm SM} = (4.2\pm 1.4) \cdot 10^{-3} }. 
\end{eqnarray}
Here the numerical values correspond to our results from parameter set I
in \eqsto{findg}{finphi}.  
In the case of $ a_{\rm fs}^s $ there is a major difference to the SM
case of Sect.~\ref{sect:sm}, which only involves $\imag
(\Gamma_{12}^s/M_{12}^s)$: in the presence of new physics $a_{\rm fs}^s$
is dominated by $\real (\Gamma_{12}^s/M_{12}^s)$ as long as
$|\phi^\Delta_s|>\phi_s^{\rm SM}$.  Thus the prediction in \eq{boundafs}
profits from the improvements due to our new operator basis --- just as
the prediction of $\dg_s$ in \eq{bounddgdm}.  From \eq{boundafs} one
also verifies the enormous sensitivity of $ a_{\rm fs}^s $ to new
physics, since it exceeds its SM value by a factor of 250 for
$\phi^\Delta_s= \pi/2$. We have plotted $a_{\rm fs}^s$ vs.\ 
$\phi^\Delta_s$ for the old and the new bases in
\fig{plotafs}.

\begin{nfigure}{tb}
\includegraphics[width=0.9\textwidth,angle=0]{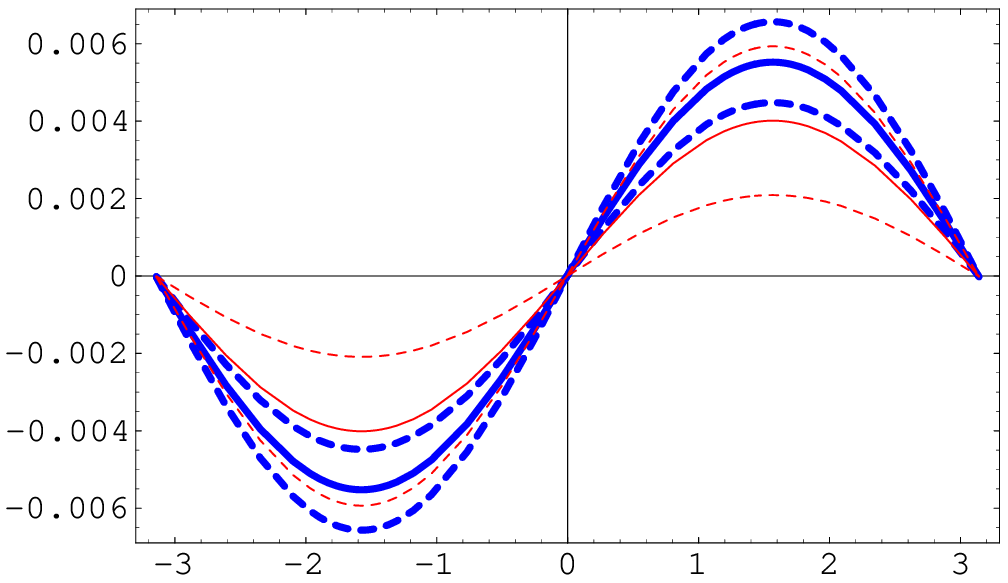}
\caption{$a_{\rm fs}^s$ as a function of the new phase $\phi^\Delta_s$
  from  \eq{boundafs} for the range 
  $-\pi \leq \phi^\Delta_s \leq \pi$. The thick blue lines show the
  prediction in the new basis, while thin red lines correspond to the 
  old operator basis.  The solid lines display the central values of our
  predictions and the dashed lines show the uncertainties, which are
  much larger for the old result. The standard model value 
  $a_{\rm fs}^s (\phi^\Delta_s=0) =  
  2.1 \cdot 10^{-5}$ is too close to zero to be 
  visible in the plot. }\label{plotafs}
\end{nfigure}

\subsection{Basic observables}\label{sect:phen}
In this section we summarise the observables which constrain 
$|\Delta_s|$ and $\phi^\Delta_s$. These constraints are illustrated in
\fig{boundband} for hypothetical measurements. 

{\bf 1.}  The mass difference $\dm_s$ determines $|\Delta_s|$ through
\eq{bounddm}.  The accuracy of $|\Delta_s|$ extracted from $\dm_s$ is
limited by the precision of a lattice computation. This is not the case
for the other quantities discussed in this section. 

Alternatively one can confront the experimental ratio $\dm_d/\dm_s$ with
theory. This has the advantage that the ratio of the hadronic matrix elements
involved can be predicted with a smaller error, of order 5\%. However, then
the parameter of $R_t$ of the unitarity triangle entering $\dm_d$ must be
taken from measurements which are insensitive to new physics (or at least
insensitive to new physics in \bbms), e.g.\ through determinations of the CKM
angle $\gamma$ from tree-level B decays (cf.\ the discussion after \eq{afsg}).
At present this method leads to comparable uncertainties in the extracted
$|\Delta_s|$ as the direct determination from $\dm_s$. (Further flavour-blind
new physics cancels from $\dm_d/\dm_s$.) In the following analyses we do not
use $\dm_d/\dm_s$.

{\bf 2.} The lifetime measurement in an untagged $b\to c\ov c s$ decay
$\BsorBsbar \to f_{CP}$, where $f_{CP}$ is a CP eigenstate, determines
$\dg_s \cos (\phi_s^\Delta-2\beta_s)=
|\dg_s \cos (\phi_s^\Delta-2\beta_s)|$ \cite{g,dfn}.
Consider a CP-even final state $f_{CP+}$ like $D_s^+D_s^-$. The
time-dependent decay rate reads
\begin{eqnarray}
\Gamma[ \BsorBsbar \to f_{CP+},t ] \propto 
        \frac{1+\cos(\phi_s^\Delta -2 \beta_s) }{2} e^{-\Gamma_L t} 
    \; +\: 
         \frac{1-\cos(\phi_s^\Delta -2 \beta_s) }{2} e^{-\Gamma_H t} \nn
\; =\; e^{-\Gamma_s t} \lt[ \cosh \frac{\dg_s \, t}{2} - 
       \cos(\phi_s^\Delta -2 \beta_s) \sinh \frac{\dg_s \, t}{2}   
       \rt] 
\label{twoexp}
\end{eqnarray}
and the (time-independent) overall normalisation is related to the 
branching fraction \cite{dfn}. Here
\begin{eqnarray}
\beta_s & = & - \arg \lt( - \frac{\lambda_t^s}{\lambda_c^s}\rt) 
   \; =\; 0.020 \pm 0.005 \; = \; 1.1^\circ \pm 0.3^\circ . \label{defbetas}
\end{eqnarray}
That is, $-\beta_s$ is the analogue of the angle $\beta$ of the unitarity
triangle, which governs the mixing-induced CP asymmetry in $B_d\to J/\psi
K_S$, in the $B_s$ system. For $\beta_s$ different sign conventions are used
in the literature, we chose the one of \cite{run2} which satisfies
$\beta_s>0$.

For example within the Standard Model (and neglecting the tiny
$\beta_s$) the lifetime measured in $\BsorBsbar \to D_s^+ D_s^-$ equals
$\Gamma^s_L=\Gamma_s+\dg/2$, because only the short-lived CP-even mass
eigenstate $B_L$ can decay into $D_s^+ D_s^-$. By using the theory
relation $1/\tau_{B_d}=\Gamma_d=(1.00 \pm 0.01) \Gamma_s$ one then finds
$\dg_s$.  For $\phi_s^\Delta\neq0$, however, the mass eigenstates are no
more CP eigenstates and both of them can decay to a CP eigenstate, as
can be easily verified from \eq{twoexp}. From $\Gamma[ \BsorBsbar \to
f_{CP+},t ]$ one can extract $|\Gamma_s|$, $|\dg_s|$, $|\cos(\phi_s^\Delta)|$
and the overall normalisation, if the statistics is high enough to 
separate the two exponentials.    
If the measured $\Gamma[ \BsorBsbar \to f_{CP+},t ]$ is fitted to a single 
exponential $\exp[-\Gamma_f \, t]$, the measured rate is \cite{hm,dfn}
\begin{eqnarray}
\Gamma_f &=& \frac{(1+\cos (\phi_s^\Delta-2\beta_s))/\Gamma_L + 
                   (1-\cos (\phi_s^\Delta-2\beta_s))/\Gamma_H}
                  {(1+\cos (\phi_s^\Delta-2\beta_s))/\Gamma_L^2 + 
                   (1-\cos (\phi_s^\Delta-2\beta_s))/\Gamma_H^2} \label{cpl1}\\
&=& \Gamma_s \; +\; \dg_s \cos (\phi_s^\Delta-2\beta_s) \; +\: 
    {\cal O} \lt( \frac{(\dg_s)^2}{\Gamma_s} \rt) \nn 
&=&
\Gamma_s \; +\; 2\, |\Gamma_{12}^s|  
       \cos (\phi_s^\Delta+\phi_s^{\rm SM}) 
       \cos (\phi_s^\Delta-2\beta_s) \; +\: 
    {\cal O} \lt( \frac{(\dg_s)^2}{\Gamma_s} \rt) . \label{cpl2} 
\end{eqnarray}
For a CP-odd final state one has to interchange $\Gamma_L$ and
$\Gamma_H$ in \eqsand{twoexp}{cpl1} and to flip the sign of $
\cos (\phi_s^\Delta-2\beta_s) $ in \eqsand{twoexp}{cpl2}. 
From \eq{cpl2} it is clear that the lifetime measurement 
determines \cite{g,dfn}
\begin{eqnarray}
\dg_s \cos (\phi_s^\Delta) &=& 2 |\Gamma_{12}^s| \cos^2 (\phi_s^\Delta), \no
\end{eqnarray}
if the small phases $\phi_s^{\rm SM}$ and $\beta_s$ are neglected.  Thus
one can find $|\cos \phi_s^\Delta|$, which determines $\phi_s^\Delta$
with a four-fold ambiguity.\footnote{If one keeps $\phi_s^{\rm SM}$ and
  $\beta_s$ non-zero, one solution for $\phi_s^\Delta$ is related to the
  other three by $\phi_s^\Delta\to \phi_s^\Delta + \pi$,
  $\phi_s^\Delta\to 2\beta_s - \phi_s^{\rm SM} -\phi_s^\Delta $ and
  $\phi_s^\Delta\to 2\beta_s - \phi_s^{\rm SM} -\phi_s^\Delta + \pi$.}
We stress that (since $\sgn\dg_s=\sgn \cos (\phi_s^\Delta)$) the
lifetime method gives no information on the sign of $\dg_s$ and
experimental results should be quoted for $|\dg_s|$ rather than $\dg_s$.

\eq{cpl1} assumes that detection efficiencies are constant over the
decay time. Since this is not the case in real experiments, we strongly
recommend to perform a three-parameter fit to $2 |\Gamma_{12}|$,
$|\cos(\phi_s^\Delta)|$ and the overall normalisation (with $|\Gamma_s|$
fixed to $|\Gamma_d|(1.00\pm 0.01)$) to \eq{twoexp}. 

With the advent of the precise measurement of $\dm_s$ \cite{dmdiscovery}
one will rather exploit $|\dg_s|/\dm_s$ to constrain $\Delta_s$ than
$|\dg_s|$ itself, which suffers from much larger hadronic uncertainties.
From \eq{bounddgdm} one infers that $|\dg_s|/\dm_s$ defines two circles in
the complex $\Delta_s$ plane which touch the y--axis at the origin.

{\bf 3.} The angular analysis of an untagged $b\to c\ov c s$ decay
$\BsorBsbar \to VV^\prime$, where $VV^\prime$ is a superposition of CP
eigenstates with vector mesons $V,V^\prime$, not only determines $\dg_s
\cos (\phi_s^\Delta -2\beta_s)$, but also contains information on $\sin
(\phi_s^\Delta -2\beta_s)$ through a CP-odd interference term.  Here
the golden mode is certainly $\BsorBsbar \to J/\psi \phi$, but also
final states with higher $\psi$ resonances and $\BsorBsbar \to D_s^{*+}
D_s^{*-}$ can be studied. The determination of $\phi_s^\Delta$ 
from the CP-odd interference term in untagged samples involves
a four-fold ambiguity. It could be reduced to a two-fold ambiguity if
the signs of $\cos \delta_1$ and $\cos \delta_2$ were determined, where
$\delta_1$ and $\delta_2$ are the strong phases involved
\cite{ddlr,dfn}. These two solution are related by $\phi_s^\Delta
\leftrightarrow \phi_s^\Delta \pm \pi $. If one relaxes the assumptions
on $\cos \delta_1$ and $\cos \delta_2$, one is back to the same 
four-fold ambiguity as in item 2. 

{\bf 4.}~ The branching fraction $Br(\BsorBsbar\to D_s^{(*)+}
D_s^{(*)-})$ approximates the width difference $\Delta \Gamma_{\rm CP}$
between the two CP eigenstates of the $B_s$ system \cite{dfn}.
Irrespective of any new physics in $M_{12}^s$ one always has $\Delta
\Gamma_{\rm CP}=2|\Gamma_{12}^s|$, so no constraint on our new physics
parameter $\Delta_s$ is gained. Yet the ratio of $\dg_s
\cos(\phi_s^\Delta -2\beta_s) $ and $\Delta \Gamma_{\rm CP}$ could
cleanly determine $\cos(\phi_s^\Delta) \cos(\phi_s^\Delta -2\beta_s) $.
However, $Br(\BsorBsbar\to D_s^{(*)+} D_s^{(*)-})$ only equals $\Delta
\Gamma_{\rm CP}$ in the poorly tested simultaneous limit of an
infinitely heavy charm quark with small-velocity \cite{sv} and an
infinite number of colours \cite{ayopr}.  In order to test this limit
one needs to measure the CP-odd and CP-even fractions of all $b\to c\ov
c s$ decays \cite{dfn}. Until this has been done nothing can be inferred
from $Br(\BsorBsbar\to D_s^{(*)+} D_s^{(*)-})$, in particular this
quantity neither gives an upper bound (since other CP-even $b\to c\ov c
s$ modes can be relevant) nor a lower bound (since other CP-odd $b\to
c\ov c s$ modes can be relevant and the $D_s^{(*)+} D_s^{(*)-}$ final
  state has a CP-odd component) on $\Delta \Gamma_{\rm CP}$. We strongly 
discourage from the inclusion of $Br(\BsorBsbar\to D_s^{(*)+}
D_s^{(*)-})$ in averages with $\dg_s$ determined from clean methods.
 
{\bf 5.} $a_{\rm fs}^s$ can be measured from untagged flavour-specific 
$\BsorBsbar$ decays, typically from the number of positively and
negatively charged leptons in semileptonic decays. Observing further 
the time evolution of these untagged 
$\BsorBsbar \to X^\mp \ell^\pm \nuornubar_\ell$ 
decays (see e.g.\ \cite{n}),
\begin{eqnarray}
\frac{ \Gamma[ \BsorBsbar \to X^- \ell^+\nu_\ell ,t ] \, -\, 
       \Gamma[ \BsorBsbar \to X^+ \ell^- \ov \nu_\ell ,t ]}{
       \Gamma[ \BsorBsbar \to X^- \ell^+\nu_\ell ,t ] \, +\,
       \Gamma[ \BsorBsbar \to X^+ \ell^- \ov \nu_\ell ,t ]}
&=& \frac{a_{\rm fs}^s}{2} \lt[ 1- 
    \frac{\cos(\dm_s \, t)}{\cosh{(\dg_s\, t/2)}}  \rt],
\label{afst}
\end{eqnarray}
will have two advantages: one can use the oscillatory behaviour to
control fake effects from experimental detection asymmetries (which are
constant in time) and to separate the $B_s$ and $B_d$ samples through
$\dm_s\neq \dm_d$. The constraint from $a_{\rm fs}^s$ on $\Delta_s$ is
given in \eq{boundafs}. It defines a circle in the complex $\Delta_s$
plane which touches the x--axis at the origin. The constraint from
$a_{\rm fs}^s$ on $\Delta_s$ only has a two-fold ambiguity (related to
$\phi_s\leftrightarrow \pi-\phi_s$) and discriminates between the solutions
in the upper and lower half--plane in \fig{boundband}.

{\bf 6.} The time dependence of the tagged decay $B_s\to J/\psi\phi $ permits
the determination of the mixing-induced CP asymmetries $A_{\rm CP}^{\rm
  mix}(B_s\to (J/\psi \phi)_{CP\pm})$. The angular analysis separates
the CP--odd P-wave component from the CP--even S-wave and D-wave.
The time-dependent CP asymmetry is (in the notation of \cite{run2,dfn}):
\begin{eqnarray}
\frac{\Gamma(\bar B^0_s (t) \to f) - \Gamma (B^0_s (t) \to f)}
     {\Gamma (\bar B^0_s (t) \to f) + \Gamma (B^0_s (t) \to f)}
& = & 
- \frac{A_{CP}^{\rm mix} \sin (\dm_s t) }
       {\cosh (\Delta \Gamma_s t/2) + A_{\Delta \Gamma} 
\sinh (\Delta \Gamma_s t/2)}.
\end{eqnarray}
One finds $\phi_s^{\Delta}-2 \beta_s $ through
\begin{eqnarray}
A_{\rm CP}^{\rm mix}(B_s\to (J/\psi \phi)_{CP\pm}) &=& 
   \pm \sin(\phi_s^{\Delta}-2 \beta_s ), \qquad 
A_{\Delta \Gamma} \; = \; \mp \cos(\phi_s^{\Delta}-2 \beta_s ) 
\label{eqamix}
\end{eqnarray}
with the same two-fold ambiguity as from $a_{\rm fs}^s$ in item 5.
Combining \eqsand{bounddm}{boundafs} with \eq{eqamix} and neglecting
the tiny contributions of $\phi_s^{\rm SM}$ and $\beta_s$ one verifies the
correlation between $a_{\rm fs}^s$ and 
$ A_{\rm CP}^{\rm mix} (B_s\to (J/\psi \phi)_{CP\pm})$ derived in 
\cite{dmphenburas, dmphenligeti}. In fact such correlations can be
found between any three of the observables discussed above, because 
the \bbms\ only involves the two parameters $|\Delta _s|$ and $\phi_s$.
 
An important remark here concerns the decay $B_s \to K^+ K^-$, as one might be
tempted to use the lifetime measured in $B_s \to K^+ K^-$ to determine
$\Gamma_s+|\dg_s/2|$. While $K^+ K^-$ is CP even, the decay is
penguin--dominated and as such sensitive to the same kind of new physics which
may be responsible for the experimental anomaly seen in penguin--dominated
$B_d$ decays \cite{ichep06}. Thus information from $B_s \to K^+ K^-$ should
under no circumstances be included in any averages with the measurements 
discussed above. Instead one should confront the lifetime measured in
this mode with the one obtained from $B_s\to (J/\psi \phi)_{CP+}$ to 
probe new physics in $b\to s$ penguin decays.


For a visualisation of the bounds from \eqsto{bounddm}{boundafs} in the
complex $\Delta_s$-plane we consider now the hypothetical case of
$|\Delta_s| = 0.9$ and $\phi^\Delta_s = - \pi / 4$.  Suppose one would
measure these central values:
\begin{eqnarray}
\Delta M_s = 17.4 \, \mbox{ps}^{-1},
&& \qquad
\Delta \Gamma_s = 0.068 \, \mbox{ps}^{-1},
\\
\frac{\Delta \Gamma_s}{\Delta M_s} = 3.91 \cdot 10^{-3},
&&\qquad
a_{\rm fs}^s = -3.89 \cdot 10^{-3} \, .
\end{eqnarray}
Moreover we assume  the following theoretical and experimental uncertainties: 
$\Delta M_s : \pm 15 \%$, 
$\Delta \phi_s : \pm 20 \%$, 
$\Delta \Gamma_s / \Delta M_s : \pm 15 \%$,
$a_{\rm fs}^s : \pm 20 \%$.
The regions in the $\Delta_s$-plane bounded for these hypothetical
measurements are shown in figure \ref{boundband}.

\begin{nfigure}{tb}
\includegraphics[width=0.9\textwidth,angle=0]{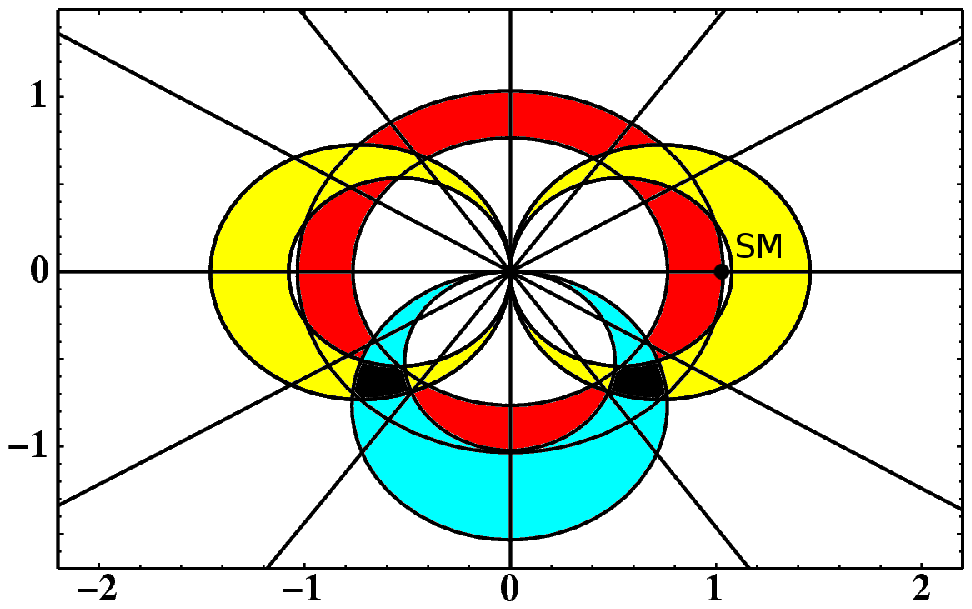}
\caption{Illustration of the bounds in the complex $\Delta_s$-plane 
  for $|\Delta_s| = 0.9$ and $\phi^\Delta_s = - \pi / 4$. 
  We assume the following overall uncertainties: 
  $\Delta M_s$                   (red or dark-grey)    : $\pm 15 \%$, 
  $\Delta \Gamma_s / \Delta M_s$ (yellow or light-grey): $\pm 15 \%$, 
  $a_{\rm fs}^s$                     (light-blue or grey)  : $\pm 20 \%$ and
  $\phi_s^\Delta$                (solid lines)         : $\pm 20 \%$.}
  \label{boundband}
\end{nfigure}
The constraints from CP-conserving quantities are symmetric to the
Im($\Delta_s$)-axis, The bound from $\Delta M_s$ simply gives a circle
with the origin (0,0) and the radius $|\Delta_s|$. In the measurement of
$\dg_s$ we have assumed that the data are fitted to the correct formula
\eq{twoexp} and $|\dg_s|$ and $|\cos (\phi_s -2\beta_s)|$ have been
determined as discussed above in item 2. In practice the extracted
$|\dg_s|$ and $|\cos (\phi^\Delta_s-2 \beta_s)|$ are strongly correlated
and mainly $|\dg_s| |\cos (\phi^\Delta_s -2 \beta_s)|$ is determined
(see \eq{cpl2} and \cite{dfn}). The constraint from the hadronically
cleaner ratio $|\dg_s|/\dm_s$ are two circles which touch the y--axis in
the origin.  If one fully includes the correlation between $|\dg_s|$ and
$|\cos (\phi^\Delta_s-2 \beta_s)|$ one will rather find constraints
which roughly correspond to a fixed $|\dg_s\cos (\phi^\Delta_s -2
\beta_s) |/\dm_s$. The corresponding curves are a bit more eccentric
than the circles from $|\dg_s|/\dm_s$.

If one plots the bounds from $|\dg_s|$ (or $|\dg_s \cos (\phi^\Delta_s
-2 \beta_s)| $) alone, one finds four rays starting from the origin. The
experimental information in this is redundant, as it is fully contained
in the constraints from $\dm_s$ and $|\dg_s|/\dm_s$. For the theory
uncertainties, however, this is not true: if (as current data do)
$\dm_s$ prefers a small value of $f_{B_s}$, while $\dg_s$ prefers a
large $f_{B_s}$, the combined constraint from $\dm_s$ and $|\dg_s|$ will
exclude a region of the $\Delta_s$ plane which is allowed by the ratio
$|\dg_s|/\dm_s$, from which $f_{B_s}$ drops out.

The measurement of $a_{\rm fs}^s$ yields a circle touching the x--axis
in the origin, in particular it reduces the four-fold ambiguity in the
extracted value of $\Delta_s$ to a two-fold one. The extraction of
$\phi^\Delta_s -2 \beta_s$ from the angular analysis in $\BsorBsbar \to
J/\psi \phi$ (as discussed in item 3) also yields four rays starting
from the origin (corresponding to the same value of $|\cos
(\phi^\Delta_s -2 \beta_s)| $), if no assumptions on the signs of
$\cos\delta_1$ and $\cos \delta_2$ are made. Finally, the measurement
of $A_{\rm CP}^{\rm mix}(B_s\to (J/\psi \phi)_{CP\pm})$ will select two
out of these four rays, discriminating between $\phi_s^\Delta -2
\beta_s>0$ and $\phi_s^\Delta -2 \beta_s < 0$.

\boldmath
\subsection{Current experimental constraints on $\Delta_s$}
\unboldmath%
In this section we turn to the real world and discuss the current
experimental constraints on the complex $\Delta_s$-plane. In view of 
the experimental errors we set $\beta_s$ to zero and identify 
$\phi_s$ with $\phi^\Delta_s$.

The mass difference $\Delta M_s$ is now known very precisely
\cite{dmdiscovery}, see \eq{dmexp}.
For the remaining mixing parameters in the $B_s$-system
only weak experimental constraints are available.
The only available experimental analysis of $|\dg_s|$ with the correct
implementation of the phase $\phi_s$ is from the D\O\ collaboration, their
analysis in \cite{dgexp2} was recently updated in \cite{dgexpnew} using
1fb$^{-1}$ of data.  Setting the value of the mixing phase $\phi_s$ to
zero (Standard Model scenario) they obtain \cite{dgexpnew}
\begin{eqnarray}
\dg_s & = & 0.12 \pm 0.08{}_{\mbox{\scriptsize (stat)}} \;   
           \epm{0.03}{0.04} \, {}_{\mbox{\scriptsize (syst)}}   
          \;  \mbox{ps}^{-1} \, .
\label{eqdmexp3}
\end{eqnarray}
Allowing for a non-zero value of the mixing phase $\phi_s$ they get
\begin{eqnarray}
\Delta \Gamma_s & = & \phantom{-} 
       0.17 \pm 0.09{}_{\mbox{\scriptsize (stat)}} \; 
          \pm 0.03{}_{\mbox{\scriptsize (syst)}}  \; \mbox{ps}^{-1} \, 
\nn &&
\mbox{and} \quad
\phi_s \; = \; -0.79  \pm 0.56{}_{\mbox{\scriptsize (stat)}}  
           \pm 0.01{}_{\mbox{\scriptsize (syst)}}  \label{eqdgexp5}\\[2mm]
\mbox{or}\qquad 
\Delta \Gamma_s & = & - 0.17 \pm 0.09{}_{\mbox{\scriptsize (stat)}}  
    \pm 0.03{}_{\mbox{\scriptsize (syst)}}  \, \mbox{ps}^{-1} \, 
\nn &&
\mbox{and} \quad
\phi_s \; = \; -0.79  \pm 0.56{}_{\mbox{\scriptsize (stat)}}  
           \pm 0.01 {}_{\mbox{\scriptsize (syst)}} \; +\; \pi  \, .
\label{eqdgexp4}
\end{eqnarray}
As expected from \eq{cpl2} the values for $|\dg_s \cos \phi_s|$ found from
\eqsand{eqdmexp3}{eqdgexp4} are roughly equal to $\dg_s$ in \eq{eqdmexp3}.
The quoted results in \eqsand{eqdgexp5}{eqdgexp4} assume that the signs of
$\cos\delta_1$ and $\cos \delta_2$ agree with the results found with naive
factorisation. With this assumption the other two solutions for $\phi_s$
(which have opposite signs to those in \eqsand{eqdgexp5}{eqdgexp4}) are
excluded. Strategies to check this theoretical input are discussed in
\cite{dfn}.

The semileptonic CP asymmetry $a_{\rm sl}^s\equiv a_{\rm fs}^s$ in the
$B_s$ system has been determined directly in \cite{aslsexp} and was found to
be
\begin{equation}
a_{\rm sl}^{s, \rm direct} = \left(24.5 \pm 19.3 
               {}_{\mbox{\scriptsize (stat)}}\pm 
      3.5 {}_{\mbox{\scriptsize (syst)}}\right) \cdot 10^{-3} \, .
\label{eqaslexp1}
\end{equation}
Moreover the semileptonic CP asymmetry can be extracted from the
same sign dimuon asymmetry that was measured in \cite{dimuonexp} as 
\begin{equation}
a_{\rm sl} = \left( -5.3 \pm 2.5 {}_{\mbox{\scriptsize (stat)}}
                         \pm 1.8 {}_{\mbox{\scriptsize (syst)}} 
             \right) \cdot 10^{-3} \, 
\label{eqdimuexp}
\end{equation}
in a data sample containing both $B_d$ and $B_s$ mesons. While the
composition of the sample is known, no determination of the initial
state on an event--by--event basis was possible.  Updating the numbers
in \cite{combined,nir2006} one sees that the measurement in
\eq{eqdimuexp} determines the combination
\begin{equation}
a_{\rm sl} = \left(0.582 \pm 0.030 \right) \, a_{\rm sl}^d + 
         \left(0.418 \pm 0.047 \right) \, a_{\rm sl}^s .
\label{asllk}
\end{equation}
In \cite{combined, nir2006} the experimental bound for $a_{\rm sl}^d$ from B
factories was used to extract a bound on $a_{\rm sl}^s$ from
Eq.(\ref{eqdimuexp}) and Eq.(\ref{asllk}). The huge experimental uncertainty
in $a_{\rm sl}^d$ then inflicts a large error on the value of $a_{\rm sl}^s$
inferred from \eqsand{eqdimuexp}{asllk}.

Here we pursue a different strategy and use the much more precise
theoretical Standard Model value for $a_{\rm sl}^d$ in \eq{afsnum}. In the
search for new physics this is permissible: if the resulting constraint 
on $\Delta_s$ departs from the Standard Model value $\Delta_s=1$, this
will then imply new physics in either $a_{\rm sl}^s$ or $a_{\rm sl}^d$. 
Moreover, the current precision in the unitarity triangle already 
substantially limits the room for new physics in $a_{\rm sl}^d$ 
\cite{exput}.   

Using $a_{\rm sl}^d = - \left( 0.48 \epm{0.10}{0.12} \right) \cdot
10^{-3}$ of \eq{afsnum} and further \eqsand{eqdimuexp}{asllk} we obtain
the nice bound
\begin{equation}
a_{\rm sl}^{s, \rm dimuon} = \left(- 12.0 
      \pm 6.0{}_{\mbox{\scriptsize (stat)}} 
      \pm 4.5 {}_{\mbox{\scriptsize (syst)}}\right) 
                             \cdot 10^{-3} \, . \label{eqasldimuon}
\end{equation}
Combining this number with the one from the direct determination
\cite{aslsexp} in \eq{eqaslexp1} we get our final experimental number
for the semileptonic CP asymmetry:
\begin{equation}
a_{\rm sl}^{s} = \left(- 8.8 \pm 5.7{}_{\mbox{\scriptsize (stat)}} 
                             \pm 4.5{}_{\mbox{\scriptsize (syst)}} 
                 \right) \cdot 10^{-3} \, .
\label{eqasldfinal}
\end{equation} 
Adding statistical and systematic error in quadrature gives 
\begin{equation}
a_{\rm sl}^{s} = \left(- 8.8 \pm 7.3 
                 \right) \cdot 10^{-3} \, .
\label{eqasldfinal2}
\end{equation} 
In Fig.~(\ref{boundbandreal}) we display all bounds in the complex
$\Delta_s$-plane including all experimental and theoretical
uncertainties.


\begin{nfigure}{tb}
\includegraphics[width=0.9\textwidth,angle=0]{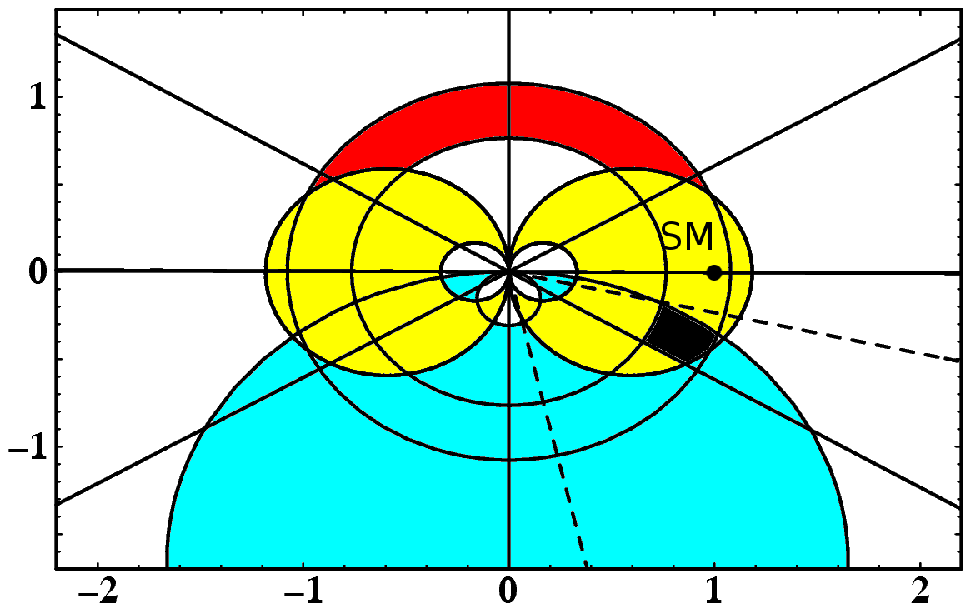}
\caption{Current experimental bounds in the complex $\Delta_s$-plane.
  The bound from $\Delta M_s$ is given by the red (dark-grey) annulus
  around the origin. The bound from $|\Delta \Gamma_s|/ \Delta M_s$ is
  given by the yellow (light-grey) region and the bound from $a_{\rm
    fs}^s$ is given by the light-blue (grey) region. The angle
  $\phi_s^\Delta$ can be extracted from $|\Delta \Gamma_s|$ (solid
  lines) with a four--fold ambiguity --- each of the four regions is
  bounded by a solid ray and the x-axis --- or from the angular analysis
  in $B_s \to J / \Psi \phi$ (dashed line). This constraint also has a
  four--fold ambiguity if no assumptions on the strong phases $\delta_1$
  and $\delta_2$ are made. The dashed lines limit the region
  corresponding to the solution in \eq{eqdgexp5}.  
  The Standard Model case corresponds to $\Delta_s= 1$.
  The current experimental situation shows
  a small deviation, which may become significant, if the experimental
  uncertainties in $\Delta \Gamma_s$, $a_{\rm sl}^s$ and $\phi_s$ will
  go down in near future.}\label{boundbandreal}
\end{nfigure}

The combined analysis of $\Delta M_s$, $\phi_s$, $|\dg_s|/\dm_s$ and
$a_{\rm sl}^s$ in Fig.~\ref{boundbandreal} shows some hints for
deviations from the Standard Model. To analyse them further we ignore 
discrete ambiguities and focus on the solution in the fourth quadrant 
which is closest to the Standard Model solution $\Delta_s=1$.
We further do not perform a complete statistical analysis with proper 
inclusion of all correlations and for simplicity add statistical and
systematic errors in quadrature. 
First we note from \eq{bounddm} that \eq{dmexp} implies 
\begin{eqnarray}  
  |\Delta_s| &=& 0.92 \pm 0.32{}_{\mbox{\scriptsize (th)}}
                \pm 0.01 {}_{\mbox{\scriptsize (exp)}} \label{resde}
\end{eqnarray}
while \eqsand{bounddgdm}{eqdgexp5} lead to 
\begin{eqnarray}  
  \frac{\cos \phi_s}{|\Delta_s|} &=&  1.93 \pm 
     0.37{}_{\mbox{\scriptsize (th)}}  \pm 1.1 {}_{\mbox{\scriptsize (exp)}} 
  \label{rescd} .
\end{eqnarray}
\eqsand{resde}{rescd} are consistent with $\Delta_s=1$, but prefer 
$|\Delta_s|<1$. 

Second we observe that both the angular distribution 
in $\BsorBsbar \to J/\psi \phi$ giving 
\eq{eqdgexp4} and $a_{\rm sl}^s$ in \eq{eqasldfinal2}  point
towards a non-zero $\phi_s$. Both analyses involve
$\sin \phi_s$, the two values inferred are 
\begin{eqnarray}
\sin \phi_s &=& -0.71 \epm{0.48}{0.27} \qquad\qquad\quad 
   \mbox{from the angular analysis, \eq{eqdgexp4},}   \label{sphia}\\
\frac{\sin \phi_s}{|\Delta_s|} &=& 
-1.77 \pm 0.33 {}_{\mbox{\scriptsize (th)}}
      \pm 1.47 {}_{\mbox{\scriptsize (exp)}} \qquad\qquad
   \mbox{from $a_{\rm sl}$ in \eq{eqasldfinal2}} .   \label{sphiasl}
\end{eqnarray} 
In \eq{sphiasl} we have profited from our improved theory prediction in 
\eq{boundafs}. For $|\Delta_s|=1$ the two numbers combine to 
\begin{eqnarray}
\sin \phi_s &=& -0.77 \pm 0.02{}_{\mbox{\scriptsize (th)}} 
                \pm 0.36 {}_{\mbox{\scriptsize (exp)}} .  \label{ressphi}
\end{eqnarray}
Relaxing $|\Delta_s|$ to its minimal
value allowed by \eq{resde}, $|\Delta_s|=0.59$, changes this
result to 
\begin{eqnarray}
\sin \phi_s &=& -0.76 \pm 0.03{}_{\mbox{\scriptsize (th)}} 
                \pm 0.34 {}_{\mbox{\scriptsize (exp)}} .  
\label{ressphi2} 
\end{eqnarray}
Either \eq{ressphi} or \eq{ressphi2} alone imply a deviation from
$\phi_s =0$ by 2.1$\sigma$, but $\dg_s$ in \eq{eqdgexp4} pulls in the
opposite direction, preferring large values of $|\cos\phi_s|$ through
\eq{bounddg}. Despite of its large error $\dg_s$ already gives a
powerful lower bound $|\cos\phi_s| \geq 0.55$ (so that $|\sin\phi_s|\leq
0.84$) at the 1$\sigma$ level because of its large central value in
\eq{eqdgexp4}. This can be clearly seen from \fig{boundbandreal}.
However, $\dg_s$ is consistent with $\cos\phi_s=0$ at the 1.8$\sigma$
level and clearly has no impact on the small $\phi_s$ region, which is
the relevant region to assess the significance of \eq{ressphi} in the
search for new physics.

In conclusion we find that the data are best fit for $\phi_s$ around
$-0.88$ corresponding to $\sin \phi_s =-0.77$, if $|\Delta_s|=1$. The
constraint from $|\dg_s|$ is less compelling, but slightly prefers
$|\Delta_s| < 0$ and disfavours too large values of $|\sin\phi_s|$. The
discrepancy between data and the Standard Model is around 2$\sigma$,
which is not statistically significant yet. If our results are used to
constrain models of new physics one should bear in mind that we have
only discussed the solution in the fourth quadrant of the complex
$\Delta_s$ plane here.

\boldmath
\section{A road map for \bbs\ mixing}\label{sect:road}
\unboldmath%
Clearly the best way to establish new physics from \bbs\ mixing is a
combination of all observables following the line of
Sect.~\ref{sect:phen}.  In particular it has to be stressed that $A_{\rm
  CP}^{\rm mix}(B_s\to (J/\psi \phi)_{CP\pm})$ and $a_{\rm fs}^s$ are
not substitutes for each other, but rather give complementary
information on the complex $\Delta_s$ plane because of their different
dependence on $|M_{12}^s|$. With the new operator basis presented in
this paper it will be possible to determine $\Delta_s$ solely from
measurements which involve hadronic quantities only in numerically
sub-dominant terms. To this end any experimental progress on $|\dg_s|$,
$a_{\rm fs}^s$, the angular distributions of both untagged and tagged
$B_s \to J/\psi \phi$ decays (with the tagged analysis giving access to
$A_{\rm CP}^{\rm mix}(B_s\to (J/\psi \phi)_{CP\pm})$) and possibly of
other $b\to c\ov c s$ decays of the $B_s$ meson is highly desirable.
Regardless of whether $\sin \phi_s$ turns out to be zero or not it is
important to measure the sign of $\dg_s$. Methods for this are discussed
in \cite{dfn}. Probably the most promising way to determine $\sgn \dg_s
= \sgn \cos (\phi_s)$ is the study of $B_s \to J/\psi K^+ K^-$ with a
scan of the invariant mass of the $(K^+,K^-)$ pair around the $\phi$
peak to determine $\sgn \cos \delta_{1,2}$.

Clearly the analysis of the precise measurement of $\dm_s$ needs a better
determination of $f_{B_s}^2 B$. Since any new physics discovery from a
quantity involving lattice QCD will be met with scepticism by the scientific
community, the lattice collaborations might want to consider to switch to blind
analyses in the future. The predictions of both $\dg_s/\dm_s$ and $a_{\rm
  fs}^s$ involve the ratio $\tilde{B}_S^\prime/B$ in a numerically
sub-dominant term. It may be worthwhile to address this ratio directly in
lattice computations, because some systematic effects could drop out from the
ratio of the two matrix elements.

The quantities discussed in this paper will also profit from higher-order
calculations of the short-distance QCD parts. In particular corrections of
order $\alpha_s/m_b$ should be computed to permit a meaningful use of
$1/m_b$ bag factors computed with lattice QCD or QCD sum rules.  A further
reduction of the dependence on the renormalisation scale $\mu_1$ requires the
cumbersome calculation of ${\cal O} (\alpha_s^2)$ corrections.  Finally, the
reduction of the $1/m_b$ corrections with the help of our new operator basis
can only be fully appreciated, if the size of the $1/m_b^2$ terms is indeed
small. We have estimated these corrections and indeed found no unnatural 
enhancement over their natural size.   

\section{Summary}\label{sect:sum}
In this letter we have improved the theoretical accuracy of the mixing
quantity $\Gamma_{12}^q$, $q=d,s$, by summing the logarithmic terms
$\alpha_s^n z \ln^n z$, $z=m_c^2/m_b^2$ to all orders $n=1,2,\ldots$ and
by introducing a new operator basis, which trades the traditionally used
operator $Q_S$ of \eq{defqs} for $\widetilde Q_S$ defined in
\eq{defqst}. In the new operator basis the coefficient of the
$1/m_b$--operator $R_0$ is colour--suppressed.  We have found that all
previously noted pathologies in the sizes of the $1/m_b$ and $\alpha_s$
corrections were artifacts of the old operator basis. Still, one could
achieve the same accuracy with the use of the old basis, if one i) used
the coefficients with resummed $\ln z$ terms, ii) added the term of
order $N_c\alpha_s/m_b$ which drops from the NLO results of
\cite{bbgln1,rome03,bbln} when $\widetilde Q_S$ is eliminated for $R_0$
and iii) fully takes the numerical correlation between $B$ and $B_S$
into account. This numerical correlation stems from the smallness of the
matrix element $\langle \widetilde Q_S \rangle$. It is most easily
implemented by expressing either $B$ or $B_S$ in terms of $\widetilde
B_S$, which is essentially equivalent to our approach.

Our improvements are most relevant for $\real \Gamma_{12}^q/M_{12}^q$,
which enters both $\dg_q/\dm_q$ and new physics scenarios of $a_{\rm
  fs}^s$. In particular, hadronic quantities now appear in these
quantities in numerically sub-dominant terms only. We have then
discussed how experimental information on $|\dg_s|$, $a_{\rm fs}^s$,
$\phi_s$ from the angular distribution of $\BsorBsbar \to J/\psi \phi$
and $A_{\rm CP}^{\rm mix}(B_s\to (J/\psi \phi)_{CP\pm})$ can be
efficiently combined to constrain the complex parameter $\Delta_s$,
which quantifies new physics in \bbms.

Armed with our more precise formulae we have analysed the combined
impact of the D\O\ analyses of the dimuon asymmetry and of the angular
distribution in the decay $\BsorBsbar \to J/\psi \phi$. Here we have
assumed that $\phi_d$ is free of new physics contributions. This is
plausible in view of the constraints on $\phi_d$ from global fits to the
unitarity triangle \cite{exput}. Scanning conservatively over theory
uncertainties, we find that $\phi_s$ deviates from its Standard Model
value by 2 standard deviations.

\section*{Acknowledgements}
This paper has substantially benefited from discussions with Damir
Becirevic, Guennadi Borissov, Sandro De Cecco, Jonathan Flynn, Bruce
Hoeneisen, Heiko Lacker, Vittorio Lubicz, Alexei Pivovarov, Junko
Shigemitsu, Cecilia Tarantino, Wolfgang Wagner, Matthew Wingate,
Norikazu Yamada and Daria Zieminska.  A.L.\ thanks the University of
Karlsruhe for several invitations and U.N.\ thanks the Fermilab theory
group for hospitality.  We are grateful to Franz Stadler for preparing
the pie charts.
We thank Luca Silvestrini for pointing out our incorrect use of 
the experimental input on the dimuon asymmetry in eq. (\ref{eqdimuexp}).

This work was supported in part by the DFG grant No.~NI 1105/1--1, by the
EU Marie-Curie grant MIRG--CT--2005--029152, by the BMBF grant 05 HT6VKB
and by the EU Contract No.~MRTN-CT-2006-035482, \lq\lq FLAVIAnet''.

%
%
%
%
%
%
%
%
%

\end{document}